\documentclass[1opt]{article}
\usepackage{graphicx,amsmath,amsfonts,amssymb,dcolumn,amsthm,euscript}
\def\bea{\begin{eqnarray}}
\def\eea{\end{eqnarray}}
\def\be{\begin{equation}}
\def\ee{\end{equation}}
\usepackage{xcolor}
\usepackage{float}   

\usepackage[perpage]{footmisc}

\newcommand\nn{\nonumber} 
\newcommand{\bq}{\begin{equation}}
\newcommand\eq{\end{equation}}

\def\bar{\overline}
\newcommand\pa{\partial}

\def\hb{\hbar}

\usepackage{changepage}

\usepackage{slashed}
\usepackage{verbatim}
\usepackage{hyperref}
\usepackage{amsmath}
\usepackage{geometry}
\usepackage{pdflscape}   
\usepackage[nottoc, notlof, notlot]{tocbibind} 
\usepackage[numbers]{natbib}
\graphicspath{{./images/}}
\usepackage{mathtools}

\numberwithin{equation}{section}

\def\demi{\frac{1}{2}}

\def\Lie {\mathcal{L}}
\def\mpm {\mu^+_-}
\def\={&=&}

\begin{document}
\title{\textbf{ 
BRST BMS4 Symmetry and its Cocycles from Horizontality Conditions  
\\ 
}}
\author{\textbf{
Laurent Baulieu
and Tom~Wetzstein
 } 
 \thanks{{\ \tt     baulieu@lpthe.jussieu.fr, twetzstein@lpthe.jussieu.fr 
  }
} 
\\\\
\textit{
LPTHE, Sorbonne Universit\'e, CNRS
 } \\
\textit{  4 Place Jussieu, 75005 Paris, France  } \\
}

\def\blue{  \color{blue}}
\def\red{  \color{red}}
\def\red{  \color{black}}
\def\blue{  \color{black}}

\def\bea{\begin{eqnarray}}
\def\eea{\end{eqnarray}}
\def\t{\tau}
\def\hb{\hbar}
\def\demi{\frac{1}{2}}
\def\blue{  \color{blue}}
\def\red{  \color{red}}
\def\black{ \color{black}}
\renewcommand\t{\tau}
\def\Diff{\mathrm{Diff}}
\def \m {\mu^z_{\bar z}}
\def \mb {\mu_z^{\bar z}} 
\def \mo {\mu^z_{0}}
\def \mbo {\mu_0^{\bar z}}
\def \mob{\mu_0^{\bar{z}}}
\def \mt {\mu^z_\t}
\def \mbt {\mu^\bz_\t}
\def \mut {\mu^z_\t}
\def \mubt {\mu^\bz_\t}
\def \mot {\mu^0_\t}
\def\w{{\wedge   }}
\def\p{\pa_z}
\def\p0{\pa_0}
\def\bp{\pa_{\bar z}}
\def\pt{\pa_\t}
\def\bz{{\bar z}  }
\def\bZ{\bar{Z}}
\def\o {\omega}
\def\O {\Omega}
\def\vp {\varphi}
\def\vpb {\bar{ \vp}}
\def\Eo{ {\cal{ E}}^z}
\def\Eob{  {\cal{ E}}^\bz}
\def\E0 { \mathcal{E}^0  }
\def\Eot {  \mathcal{E}^\t }
\def\Et{\mathcal{E} }
\def\Do{ {\cal {D}}_0  }
\def\Dz{ {\cal {D}}_z  }
\def\Dbz{ {\cal {D}}_\bz  }
\def\Lie {\mathcal{L}}
\def\mmb{1-\m\mb }
\def\Dt{ {\cal {D}}_\tau  }
\def \pbM  {   \begin{pmatrix}     }
\def \peM  {   \end{pmatrix}     }
\def \bM  {   \begin{matrix}     }
\def \eM  {   \end{matrix}     }
\def \Mo {\pbM  \mo & \mt \\ \mbo & \mbt  \peM}
\def \idd {\pbM 0 & 1 \\ -1 & 0  \peM}
\def \mmbo {\mo \mbt - \mbo \mt }
\def \mmo {\mbo \mt - \mo \mbt	}
\def \mpm {\mu^+_-}
\def\DD{\mathbb{D}}
 \def \invMu { \begin{pmatrix}
1
&
-\m
 \\ 
-\mb
  &
1 
\end{pmatrix}  }
  \def \Mut
{ \begin{pmatrix}
1
&
\mb
 \cr 
\m
  &
1 
\end{pmatrix}  
}

  \def \Mu
{ \begin{pmatrix}
1
&
\m
 \cr 
\mb
  &
1 
\end{pmatrix}  
}

\date{
$ $
April 26, 2023
}

\maketitle

\hskip 8cm \textit {  $ $}

\hskip 8cm \textit{ }

\vskip 1cm

\begin{abstract}
The  BRST structure of the extended  Bondi--Metzner--Sachs symmetry group of asymptotically flat manifolds is investigated using  the recently introduced framework of the Beltrami field parametrization of four-dimensional metrics. 
The latter identifies geometrically the two  physical degrees of freedom of the graviton as fundamental fields.   The   graded BRST BMS4  nilpotent differential  operator  relies on  four horizontality conditions  giving a   Lagrangian reformulation of the asymptotic  BMS4  symmetry.  A series  of cocycles is found which  indicate the possibility of anomalies for  three-dimensional Lagrangian theories to be    built  in  the null boundaries of asymptotically flat spaces from the principle of  BRST  BMS4 invariance.

\end{abstract}
  
\newpage
\tableofcontents 
\newpage

\section{Introduction}

This paper analyses  the  BRST structure of the infinite-dimensional asymptotic  symmetry group of asymptotically flat four-dimensional   Lorentzian manifolds $\mathcal{M}_4$.  This group was  discovered in the early $60$'s by Bondi, Metzner, van der Burg \cite{bondi} and Sachs \cite{sachs} and  it  is currently denoted as the BMS4 group. It   contains  the  Poincar\'e group  as one of its  subgroups as well as   an infinite-dimensional extension of the translations, known as the supertranslations.  
It    can be further enlarged to also contain the so-called superrotations  that are an infinite-dimensional generalization of the Lorentz transformations and  play a key role  in the context of flat space holography   \cite{deBoer, BMS/CFT}. One then gets     the extended BMS4 group. 
 Bigger asymptotic symmetry groups were later on  discovered
by relaxing the   falloff conditions of  asymptotically flat spaces   and/or  softening part of the Bondi gauge conditions  as discussed in \cite{Compere, Freidel, Geiller}.

The   seminal work of  Strominger \cite{Strominger_BMS_scattering}  has inspired  deeper studies of the BMS symmetry.
In particular, \cite{BMS_soft_graviton}   exhibits   links     between the Ward identities for       the supertranslation symmetry, the Weinberg  soft graviton theorem  and the gravitational memory effects \cite{BMS_memory} for gravity.  Analogous links also exist for QED.  Such 
  relations between   three   apparently uncorrelated area of physics can be  pictured  as  ``infrared triangles" (see \cite{Strominger_lectures} for a review).  They  have been  generalized  by including  the superrotation invariance  in \cite{Barnich_superrotations,Campiglia},  that gives relations  between  the subleading soft graviton theorems    \cite{Cachazo} and a  
  description of  spin memory effects \cite{Pasterski}.
\cite{Ciambelli_lectures}  reviews these     recent  developments   that  also involve  studying    the charge algebra of the presently known  versions of the BMS symmetry within  the covariant phase space formalism \cite{Wald,Barnich_Charge_algebra}.  It is  noteworthy  that  the algebra      of these  Hamiltonian   charges  may exhibit a central extension  related to the existence of  a non trivial $2$ cocycle for a BRST BMS  Hamiltonian operator \cite{Barnich_BRST}.

   The goal of this paper is  actually to properly construct the Lagrangian  BRST  operation for the BMS4 symmetry   and  determine  the possible cocycles that  
  may control the model dependent  anomalies   of  three-dimensional  local quantum field theories constructed from  the principle of the  BRST BMS4 invariance in  the null boundaries  $\cal I^\pm$ of $\mathcal{M}_4$.
This is done within   the recently introduced   leaf of  leaf foliation  scheme \cite{leafofleaf}.   The  latter  generically   refines the   ADM    foliation \cite{ADM} and provides  a covariant and local decomposition  of the ADM leaves $\Sigma_{d-1}=\Sigma_2\times  \Sigma_{d-3}$ of Lorentzian manifolds $\mathcal{M}_d$  
by   randomly  varying surfaces $\Sigma_2 $.  This  
        description
 is  analytically  expressed  by 
 a  so-called  generalized Beltrami  parametrization of   $d$-beins in terms of $d(d+1)/2$ local fields, referred to as
 $d$-dimensional  Beltrami 
 fields,    which suggestively   generalize in  higher dimensions    the   well-established 3 Beltrami fields $\m,
 \ \mb$ and $\Phi$  that covariantly parametrize the 
  bidimensional Beltrami metric $ds^2=\exp \Phi ||dz+\m d \bz||^2$.  The   bidimensional Beltrami  parametrization   has proven to be  a  very useful tool to study many  aspects of
  $2d$ gravity and supergravity since the   $80$'s \cite{Baulieu_Marc,Baulieu_Singer,Baulieu_Stora,Stora}. 
  It is certainly rewarding that its generalization  for $d>2$ ($d=4$ in this paper) provides new  suggestive perspectives on the various metric components while  respecting the covariance principles and taking care  of the propagation of physical gravitons.
Constructing the  generalized  $ d>2$  covariant Beltrami parametrization  in \cite{leafofleaf}  was  actually  motivated
by the search for a covariant decomposition of the $d$-dimensional metric components into distinct
sectors, each of them having its physically relevant interpretation.  It was found that   the  physical $d(d-3)/2$   degrees of freedom  of    massless $d$-dimensional gravitons can be identified  within the  leaf of leaf metric decomposition  as a well-defined subset of  the 
  $d(d+1)/2$    Beltrami fields,  geometrically identified  as the fundamental  gravitational   fields.  
  
If one focuses on the null boundaries of four-dimensional asymptotically flat space,      the surfaces $\Sigma_2$  can be identified with the celestial spheres foliated along their  null directions.  The excitations of their Beltrami differential $\m$ and $\mb$ stand for the field description of the two physical helicity states of a graviton, as will be shown in great details in a different paper.    The definition of the BRST symmetry associated to    the four-dimensional BMS4 group, 
  generically denoted as the  BRST BMS4 symmetry in this paper, will be built through geometrical   horizontality conditions within the context of the  $d=4$ Beltrami parametrization. 
  
  It will   be shown  that 
BRST BMS4 descent equations 
exist  and define   cocycles,     that indicate     the existence  of 
   three-dimensional  anomalies  for theories defined at the tree level from 
  the principle of BMS4-BRST invariance in $\cal I^\pm$. Getting  what can be called the fundamental space representation of the BMS4 algebra is actually a quite transparent task from the point of view of the BRST BMS4 symmetry,  as it can be  geometrically defined.
   Classifying  all its higher space representations is a very non trivial question.

   

  
It must be noted  that the  BRST Beltrami construction  presented in this work  is    quite appropriate to     construct   supersymmetric extensions of the BMS group. Such structures have been called super-BMS group \cite{Awada_superBMS,Fuentealba}.  This could be done by extending the genuine  pure gravity horizontality conditions used in this work to  those  of topological gravity before twisting it to get $N=1$ supergravity.  The way it goes  is to be presented  in a different publication.  The Beltrami construction   also   clarifies other  aspects of the graviton emissions and absorptions near the null boundaries of spacetime,   as well as their relation with the vacuum changes of the celestial sphere, to  be also  published in another   publication. 

The rest of the paper is organized as follows. Section[\ref{Section2}] recalls the basic ingredients of the four-dimensional Beltrami parametrization and adapts it to the specific falloff conditions of asymptotically flat spacetimes. 

Section[\ref{Section3}] explicitly constructs the nilpotent BRST BMS4 operator as stemming from four-dimensional horizontality conditions. The way this operator acts on the data at null infinity and on the basic ghost fields of the theory is exposed.

  Section[\ref{Section4}] exposes a non trivial cocycle $ \tilde{\Delta}_4$, solution of the Wess and Zumino consistency conditions $(s+d)\tilde{\Delta}_4=0$ where $s$ is the local  nilpotent  differential operator that represents the  BRST BMS4 symmetry. The explicit form of $\Delta_3^1$ is discussed. Its  existence  indicates   possible model dependent  breakings of the   Ward identity of  the BRST BMS4 symmetry,  which can possibly  invalidate some of the local $3$-dimensional quantum field theories one may consider in~$\mathcal{I}^+$.
 
 Finally,   appendices  [\ref{Annexe_A}] and  [\ref{Annexe_D}]  respectively compute the  four-dimensional Spin Connection and the BRST transformations of the ten Beltrami fields, whose detailed expressions are necessary ingredients for the work presented here.

\newpage

\section{Four-dimensional Beltrami Parametrization of Asymptotically Flat Manifolds}
\label{Section2}

\subsection{Beltrami Parametrization}

\cite{leafofleaf} defines the covariant Beltrami parametrization of a generic vierbein as follows
\bea\label{Bele}
\pbM
e^z\cr e^\bz \cr e^0 \cr e^\t
\peM_{\rm Beltrami}
\equiv
\pbM
\exp \vp
&0&0&0
\cr
0&\exp \vpb &0&0
\cr
0&0&N&0\cr
0&0&0&M
\peM
  \pbM
1 &\m&
\mu^z_0
&
\mu^z_\t
\cr
\mb 
&
1
&
\mu^\bz_0
&
\mu^\bz_\t
\cr
0
&
0
&
1&
  {\mu^0_\t} 
\cr
0
&
0
&
-  {\mu^0_\t} 
&
1
\peM
\pbM
dz\cr d \bz \cr d x^0 \cr d \t
\peM,
\label{dse}
\eea
where $\vp = \vpb \equiv \frac{\Phi}{2}$.   \eqref{dse} is the generalization of the     Beltrami zweibein  
 $e^z\equiv \exp  {\frac \Phi 2 }   (dz+\m d\bz),\  e^\bz\equiv  \exp  {\frac \Phi 2 }    (d\bz+\mb dz)$ as it was defined in \cite{Baulieu_Marc} in the context of string theory.  It  provides  a covariant and local decomposition  of the ADM leaves $\Sigma_{3}=\Sigma_2\times  \Sigma_{1}$ of Lorentzian manifolds $\mathcal{M}_4$  
by   randomly  varying surfaces $\Sigma_2 $.  Here     $(z,\bz)$ are complex coordinates  for   $\Sigma_2$, $x^0\equiv t $ is for the real  spatial transverse  direction for this set of coordinates and $\t$ is  for the Lorentz time.  For asymptotically flat manifolds, $\Sigma_2$ is a Riemann sphere at spatial infinity and the Minkowski metric writes 
\begin{equation}
\label{Mink_sphere}
ds^2_{flat}  =  \eta_{ab} dx^adx^b= -d\t^2 + dt^2 + t^2 \gamma_{z\bz} dz d\bz
\end{equation}
with $\gamma_{z\bz} = \frac{4}{(1+z\bz)^2}$, so that $z$ and $\bz$ are complex coordinates on the unit sphere.
In polar coordinates $ \gamma_{z\bz} dz d\bz= d\theta^2  +\sin (\theta) ^2 d\varphi^2$. 
 The Beltrami  line element  
 is recovered by using  \eqref{Bele} and computing     $g_{\mu\nu} = e^a_\mu \eta_{ab} e^b_\nu$, giving   
\begin{equation}
\label{dsb}
 ds^2=  -  M^2 (d\t - \frac a M dt )^2 +  N^2  (dt +\frac    a N d\t)^2
  +
 t^2 \gamma_{z\bz} \exp\Phi ||dz+  \m  d\bz+\mu^z_0 dx^0 +\mu^z_\t d\t||^2,
\end{equation}
where  $a \equiv\mot$,  with a slight change of notations as compared to \cite{leafofleaf}.   

The metric 
 \eqref{dsb} or, equivalently, the vierbein  \eqref{Bele}, defines the   ten four-dimensional   ``Beltrami fields"   $M,N, \Phi,\mot, \m,\mb,\mo,
\mbo, \mu^z_\t  ,\mbt 
$  in coordinates  $(\t, t, z,\bz)$, analogously  as the   bidimensional formula  
  $ds^2=\exp \Phi  ||d z +\m d\bz||^2 $ defines    the  standard Beltrami  differential $\m$  
 and the conformal factor $\Phi$ for a Riemann surface.  The   seven fields $\mu^a_b$ are Weyl invariant
  while $M,N, \Phi$ transform non trivially under Weyl transformations.


 The   definition \eqref{Bele} is supported    by   a $\Diff_4\subset \Diff_4\times$Lorentz covariant  gauge fixing 
of   the   $16$  components  of any given   generic    vierbein    into   the ten  Beltrami fields
 that  parametrize  both the vierbein and the $d=4$ metric
  in    \eqref{Bele} and   \eqref{dsb}.   \cite{leafofleaf} indicates that these $6$ gauge functions  split into  
the $5$  ``Beltrami  conditions"   $e^0_z=e^0_\bz=e^\t_z=e^\t_\bz=0$, $ a\equiv \mu^0_\t=-\mu^\t_0$ and  a $6$th one,     imposing  that 
    $e^z$ and $e ^\bz$ have the same conformal factor $\exp \Phi$. 
    After such a Lorentz gauge fixing,  the 
Lorentz ghosts   get fixed  as  functions of the reparametrization ghosts, 
so that the  reparametrization  BRST invariance of the ``Beltrami  conditions" is warranted. 
  Appendices[\ref{Annexe_A}] and [\ref{Annexe_D}]  compute  their expressions as well as  that of  the Spin connection  $\o (e)$ that solves the condition
$T=de+\o(e)\w e=0$ and its consistency condition  $DT^a = R^{ab}\wedge e_b=0$, in function of the Beltrami fields.

 The use  of   light cone coordinates $\t^\pm = \t \pm t$ is  a must in the framework of the BMS symmetry and of  the gravitational waves.   The 
    light cone  Beltrami metric  equivalent  to \eqref{dsb} is 
   \begin{equation}
   \label{dslc}
  ds^2
=
  -  { \cal M}^2  (d\t ^+   +\mu^+_ - d\t ^-   ) (d\t ^-   +\mu^-_+      d\t ^+   ) 
    + r^2 \gamma_{z\bz} \exp \Phi  || dz+\m d\bz+\mu^z _+  d\t^+  +\mu^z _-  d\t^- ||^2
\end{equation}
and the  corresponding Beltrami   light cone vierbein 
is
\bea
{e^{a}_{\rm Beltrami} }=
\pbM  \exp \frac{\Phi}{2} & 0 & 0 & 0 \\ 0 & \exp \frac{\Phi}{2} & 0 & 0 \\  0 & 0 & \mathcal{M} & 0 \\  0 & 0 & 0 & \mathcal{M}  \peM  \pbM 1 & \m & \mu^z_+ & \mu^z_- \\ \mb & 1 & \mu^\bz_+ & \mu^\bz_- \\ 0 & 0 & 1 & \mu^+_- \\ 0 & 0 & \mu^-_+ & 1  \peM   
\pbM   dz \\ d\bz \\ d\t^+ \\ d\t^-  \peM  .
\eea
 The consistency between  the   two sets of coordinates   in     \eqref{dsb} and \eqref{dslc} implies  that the  fields $(\Phi,\m,\mb)$ remain the same when one passes  from  one system to the other and that one has 
\begin{align}
\label{equiv_beltrami}
&\mu^z_- (z,\bz,\t^\pm) = \demi ( \mt - \mo ) , &   &\mu^\bz_- (z,\bz,\t^\pm) = \demi ( \mbt - \mbo ),
\nn \\
&\mu^z_+(z,\bz,\t^\pm) = \demi ( \mt + \mo  ),  &  &\mu^\bz_+ (z,\bz,\t^\pm) = \demi ( \mbt + \mbo  ),
\nn \\
&\mpm (z,\bz,\t^\pm)= \frac{2a + M - N}{M + N} ,  &    & \mu^-_+ (z,\bz,\t^\pm) = \frac{- 2a + M - N}{M + N},
\nn \\
&\mathcal{M}^2 (z,\bz,\t^\pm) = \frac{1}{4} ( M + N  )^2 ,
\end{align}
where the fields in the r.h.s. of these equalities are   functions  of $(z,\bz,t,\t)$. 


\subsection{Bondi Gauge and Falloff of the Beltrami Fields}

For the rest of this article, attention will be focused on future null infinity $\mathcal{I}^+$, although all of the results are easily adapted to $\mathcal{I}^-$. The   Bondi metric near $\mathcal{I}^+$ in retarded    Finkelstein--Eddington 
coordinates $(u \equiv \t^-,r \equiv t,z,\bz)$ is often denoted  as   \cite{sachs}
\begin{equation}
\label{Bondi_metric}
ds^2 = e^{2 \beta} \frac{V}{r} du^2 - 2e^{2 \beta} dr du + g_{AB} (dx^A - U^A du)( dx^B - U^Bdu ) .
\end{equation}
 Here $A=z,\bz$ or $A=\theta,\phi$ refer to the  polar coordinates on the possibly distorted celestial  sphere. 
 The  so-called  Bondi gauge  functions are
 \begin{align}\label{bgauge}
g_{rr}  &= g_{rA}   =   \pa_r \left( \frac{\det g_{AB}}{r^4} \right) =0
\end{align}
 and the    falloff conditions of asymptotically flat spacetimes that solve Einstein's equations are 
\begin{align}
\label{fall-off_conditions}
g_{uu} &=  - 1 + \frac{2 m_B}{r} + \mathcal{O}(r^{-2}) ,
\nn\\ g_{ur} &= -1 + \mathcal{O}(r^{-2}) , 
\nn \\
g_{AB} &= r^2 \bar{\gamma}_{AB} + r \ C_{AB} +  \mathcal{O}(1) ,
\nn\\ g_{uA} &= \mathcal{O}(1) .
\end{align}
BMS \cite{bondi,sachs} have shown that
\eqref{Bondi_metric} and \eqref{fall-off_conditions} provide a most practical and precise   description of  gravitational plane  waves near the null boundaries of ${\cal M}_4$. 
 One commonly   calls   
  $m_B(u,z,\bz)$   the Bondi mass aspect and $C_{AB}$  the traceless shear tensor that  describes gravitational waves (the traceless property comes from the determinant gauge condition). 
   The background metric $\bar{\gamma}_{AB}$ is conformally related to the metric of the unit $2$-sphere as $\bar{\gamma}_{AB} dx^A dx^B = \gamma_{z\bz} \exp \Phi dz d\bz$.

It turns out that the Bondi gauge is  simply  recovered by imposing the three gauge conditions
\begin{equation}
\label{gcb_2}
\mu^-_+ =0, \quad  \mu^z_+ =0,  \quad  \mu^\bz_+ = 0
\end{equation}
in the Beltrami light cone metric \eqref{dslc}. Indeed, by going to the retarded    Finkelstein--Eddington  coordinates by imposing $\t^+ = u + 2r$, the metric rewrites 
\begin{equation}
\label{lcbb1}
ds^2 = - \mathcal{M}^2 (1 + \mu^+_-) du^2 -2 \mathcal{M}^2 dr du  + r^2 \gamma_{z\bz} \exp \Phi ||    dz + \m d\bz + \mu^z_- du    ||^2.
\end{equation}
The   Bondi metric \eqref{Bondi_metric}  is thus recovered in the Beltrami parametrization, according to 
\begin{align}
\label{equiv_bel_bondi}
\mu^+_- &= - 1 - \frac{V}{r}  ,
\nn\\
\mathcal{M}^2 &= e^{2 \beta} ,
\nn \\
\pbM \mu^z_- \\ \mu^\bz_- \peM  &= - \Mu  \pbM U^z   \\  U^\bz  \peM  .
\end{align}
\eqref{gcb_2} can thus be   understood  as a gauge fixing to zero  of the  shift vector  of a mini foliation, near $\cal I^+$ and  along the $d\t^+$ direction,  which is the direction of possibly    absorbed  gravitons in $\cal I^+$.
 One immediately infers that this gauge fixing of the metric  is to  be incomplete, analogous to a  Yang--Mills gauge fixing in the temporal gauge, leading one to zero modes in the reparametrization Faddeev--Popov ghosts.  Their study within the
$d=4$  Beltrami parametrization and its BRST symmetry  is actually the subject of this work \footnote{Getting a full gauge fixing involves a ghost of ghost  methodology, analogous to that one must use in unimodular gauges \cite{Baulieu_unimodular}, and it will be studied in a separate article.}.

Each  Beltrami field $X$ can be  expanded near $\mathcal{I}^+$ as a Laurent series in function of $r$ with  the  following notation
\bea 
\label{Beltrami_exp}
X   (u,r,z,\bz )\equiv  \sum_{p} {r^{-p}} {X}_{-p} (u,z,\bz).
\eea       
The first line of \eqref{equiv_bel_bondi} implies  that the term of order $1$ in the $1/r$ expansion of $\mpm$ can be identified with the Bondi mass aspect, according to
\begin{equation}
m_B(u,z,\bz) = - \demi {\mpm}_{-1}.
\end{equation}
$\m,\mb$ and $\Phi$ build      a standard   Beltrami parametrization of  the $3$ degrees of freedom of the bidimensional matrix~$g_{AB}$, such that  
\bea
g_{AB}= r^2 \gamma_{z\bz} \frac{\exp\Phi}{2}   \pbM
2 \mb & 1+\m\mb  \\
 1+\m\mb  & 2\m
 \peM  =r^2 \gamma_{z\bz}
 \Mut
 \pbM
 0&1\\
1&0
 \peM
 \Mu
\eea 
 so that one has
\begin{equation}
g_{AB} (dx^A - U^A du)( dx^B - U^Bdu ) = r^2 \gamma_{z\bz} \exp \Phi  ||dz + \m d\bz +\mu  ^z_- du ||^2 .
\end{equation}
If one uses  the  $(z,\bz,t,\t)$ system of coordinates,   \eqref{equiv_beltrami} implies that the   gauge choice    \eqref{gcb_2}   is     \begin{align}
   \label{7.14}
\mt  &= \mu^z_u,    \quad      \mo =-  \mu^z_u ,
\nn\\
\mbt &= \mu^\bz_u,     \quad    \mbo   = - \mu^\bz_u, 
\nn\\
N &= \mathcal{M}(1 - \demi \mpm )   ,
\nn \\  
M &=   \mathcal{M} ( 1 + \demi \mu^+_-)  ,
\nn \\
a &= \frac{\mathcal{M}  \mpm }{2} 
\end{align}
where $\t^- \equiv u$. The  determinant  gauge condition is 
\begin{align}
\label{constraint_phi}
  \pa_r \Phi  =  \frac{\pa_r (\m \mb)}{(\mmb)}  . 
\end{align}
At first non trivial    order $\mathcal{O}(r^{-1})$,  \eqref{constraint_phi} implies 
\begin{equation}
\label{link_Phi_m}
\Phi_{-1} = {\m}_{-1} {\mb}_0 + {\mb}_{-1} {\m}_0.
\end{equation}
This constraint  is to  play an important role  when   combined    with   the  Beltrami field falloff conditions near~$\cal I^+$.  In fact, the latter are obtained by plugging   the definition of the   Beltrami     metric     \eqref{lcbb1} 
within the   falloff conditions \eqref{fall-off_conditions}. The  asymptotic Beltrami fields are  to satisfy  the following $1/r$ expansions for large values of $r$
\begin{align}
\label{power_expansion} 
\mu^+_- &= \frac{{\mu^+_-}_{-1}(u,z,\bz)}{r} + \mathcal{O}(r^{-2})   ,
\nn \\
\mu^z_\bz &= \frac{{\mu^z_\bz}_{-1}(u,z,\bz)}{r} + \mathcal{O}(r^{-2})  ,
\nn \\
\mu^\bz_z &= \frac{{\mu^\bz_z}_{-1}(u,z,\bz)}{r} + \mathcal{O}(r^{-2}) ,
\nn \\
\mu^A_u &= \frac{{\mu^A_u}_{-2}(u,z,\bz)}{r^2} + \mathcal{O}(r^{-3})  ,
\nn \\
\mathcal{M} &= 1 +  \frac{\mathcal{M}_{-2}(u,z,\bz)}{r^2} + \mathcal{O}(r^{-3}) ,
\nn \\
\exp \Phi &= \o(u,z,\bz) + \frac{\Phi_{-2}(u,z,\bz)}{r^2}  +  \mathcal{O}(r^{-3}) .
\end{align}
$\Phi_{-1}$ is set to zero in \eqref{power_expansion} because of  \eqref{link_Phi_m} and ${\m}_0 = {\mb}_0 = 0$, due to the  falloffs of $\m$ and $\mb$.  This is a major simplification when building the BRST operator associated to the BMS symmetry.

\section{Extended BMS4 Algebra from BRST Horizontality Conditions}
\label{Section3}

The extended BMS4 symmetry is defined as the residual $\Diff_4$ symmetry that leaves invariant the Bondi gauge \eqref{gcb_2} and the falloffs \eqref{power_expansion}. The existence of the latter has been justified by the analogy between \eqref{gcb_2} and the Yang--Mills temporal gauge.  
The associated BRST symmetry operation for this restricted $\Diff_4$ symmetry   needs introducing its (Faddeev--Popov) ghosts as anticommuting vector   $\Diff_4$ reparametrization ghosts $\xi^\mu(x)$, with an  $x$  dependence  restricted  by the   four   
 consistency conditions  for the Bondi gauge  at all order in the $1/r$ expansion
\begin{align}
\label{stability_conditions}
s\mu^-_+ = 0,   \quad s\mu^z_+ =0, \quad  s\mu^\bz_+ = 0, \quad  g^{AB}  s g_{AB} =0
\end{align}
and, for the falloff conditions, by
\begin{align}
\label{s_falloffs}
s \mpm &= \mathcal{O}(r^{-1}),
\nn \\
s \m &= \mathcal{O}(r^{-1}),
\nn \\
s \mb &= \mathcal{O}(r^{-1}),
\nn \\
s \mu^A_u &= \mathcal{O}(r^{-2}),
\nn \\
s \mathcal{M} &= \mathcal{O}(r^{-2}),
\nn \\
s \Phi &= \o' (u,z,\bz) + \mathcal{O}(\ln(r^{-1})).
\end{align}
 
 \eqref{stability_conditions} can be seen as the equations of motion of the antighosts  stemming from the BRST exact terms that ensure
the 4 gauge functions \eqref{gcb_2}. 
They are the generalization of  the degeneracy of the  Yang--Mills ghost equation of motion  when one uses the  temporal gauge $A_\t=0$, which  provides 
$sA_\t =  \pa_\t c =0$. Getting degenerate Faddeev--Popov ghosts is often the consequence of using so-called ``physical gauges".

By relaxing the conditions on $s\m$ and $s\mb$, that is allowing for  non zero ${\m}_0$ and ${\mb}_0$ in the background metric $\bar{\gamma}_{AB}$ would lead to what is known as the generalized BMS4 algebra. The reasoning that follows could also be applied in this more general case, but brings no real extra information due to the zero genus of the celestial sphere.

Appendix[\ref{Annexe_D}]    computes the BRST transformations  $sX _{\rm Beltrami}$ of the ten Beltrami fields $X_{\rm Beltrami}$ from  the  ghost dependent  geometrical   torsion  free  equation that  encodes all the  $\Diff_4$  BRST  symmetry   \cite{p_forms}
\begin{equation}
\label{topological_torsion_main}
\tilde{T}^a \equiv  (d+s) \tilde e^a+(\omega^{a}_{\ b }  + \Omega^{a}_{\ b} ) \w   \tilde e^b=  \exp i_\xi \ T^a = 0
\end{equation}
where  $\xi^\mu(x) $ is the anticommuting reparametrization ghost,  $\o^{ab}(x)$ and $\O^{ab}(x)$ the Lorentz  Spin connection and ghost respectively and $\tilde e^a \equiv \exp i_\xi e^a$.  
This permits one to express the consistency conditions \eqref {stability_conditions}. 
The   found  nilpotent BRST operator $s$ acting on the  Beltrami fields is in fact equivalent  to that acting on a generic metric 
$g_{\mu \nu}$    with 
\begin{align}
s g_{\mu \nu} &= \Lie_\xi g_{\mu \nu} =  \xi^\rho \pa_\rho g_{\mu\nu}  + g_{\mu\rho } \pa_\nu \xi^\rho  + g_{\nu\rho } \pa_\mu \xi^\rho,
\nn \\
s \xi^\mu &= \xi^\alpha \pa_\alpha \xi^\mu. 
\end{align}
The  four   consistency conditions   \eqref{stability_conditions}  of the Bondi  gauge  \eqref{bgauge} are therefore  
 \begin{align}
\label{Lie_xi_g}
0 &= \Lie_\xi g_{rr} = -2 \mathcal{M}^2 \pa_r \xi^u  ,
\nn \\
0 &= \Lie_\xi g_{rz} = - \mathcal{M}^2 \pa_z \xi^u + \mb \exp { \Phi'} \  \pa_r \xi^z + \demi (1 + \m \mb) \exp { \Phi'} \  \pa_r \xi^\bz  ,
\nn \\
0 &= \Lie_\xi g_{r\bz} =  - \mathcal{M}^2 \bp \xi^u + \m \exp { \Phi'} \  \pa_r \xi^\bz + \demi (1 + \m \mb) \exp { \Phi'} \  \pa_r \xi^z  ,
\nn \\
0 &= g^{AB} \Lie_\xi g_{AB} 
= { \xi^r \left(   \frac{2}{r} + \pa_r \Phi - \frac{ \pa_r (\m \mb) }{\mmb}   \right) } +  \xi^i \left( \pa_i { \Phi'} - \frac{ \pa_i (\m \mb) }{\mmb}   \right) 
 +\left(  \pa_A \xi^A + \mu^A_u \mathcal{D}_A \xi^u \right) .
\end{align}
Here, $\exp \Phi' \equiv r^2 \gamma_{z\bz} \exp \Phi$ and   the indices  $i$  and $A$ run respectively over $ (u,z,\bz)$ and  $(z,\bz)$.    $(\mathcal{D}_z, \mathcal{D}_\bz)$ stand~for 
\begin{align}
\label{def_D_z}
\pbM  \Dz \\ \Dbz   \peM &\equiv \frac{1}{\mmb} \pbM  
\pa_z - \mb \bp
\\
\bp - \m \pa_z
\peM.
\end{align}
Using \eqref{constraint_phi}, one can solve the    equations \eqref{Lie_xi_g} that are  valid to all order in the $1/r$ expansion. It  provides the following metric dependent restrictions for $\xi^\mu(u,r,z,\bz )$:
\begin{align}
\label{ghosts_without_EoM}
\xi^u(u,r,z,\bz ) &\equiv \xi^u_0(u,z,\bz) ,
\nn \\
\xi^z(u,r,z,\bz ) &\equiv \xi^z_0(u,z,\bz) - 2 \bp \xi^u_0 \int_r^{\infty}  dr' \ \frac{ \mathcal{M}^2 (1 + \m \mb ) \exp -{ \Phi'}}{(\mmb)^2} + 4 \pa_z \xi^u_0 \int_r^\infty dr' \ \frac{\mathcal{M}^2 \m \exp - { \Phi'}}{(\mmb)^2}  ,
\nn \\
\xi^\bz(u,r,z,\bz ) &\equiv \xi^\bz_0(u,z,\bz) - 2 \pa_z \xi^u_0 \int_r^{\infty}  dr' \ \frac{ \mathcal{M}^2 (1 + \m \mb ) \exp -{ \Phi'}}{(\mmb)^2} + 4 \bp \xi^u_0 \int_r^\infty dr' \ \frac{\mathcal{M}^2 \mb \exp - { \Phi'}}{(\mmb)^2} ,
\nn \\
\xi^r(u,r,z,\bz )  &= - \frac{r}{2} \left[ \left(   \pa_A \xi^A + \mu^A_u \mathcal{D}_A \xi^u \right) + \xi^i \left( \pa_i { \Phi'} - \frac{ \pa_i(\m \mb) }{\mmb}  \right) \right] .
\end{align}
 At first non trivial order in  $1/r$  one gets
\begin{align}
\label{xi_BMS_before_s_metric}
\xi^z (u,r,z,\bz)&=  \xi^z_0 (u,z,\bz) -  \frac{1}{r} \nabla^z \xi^u   + \mathcal{O}(r^{-2}) ,
\nn \\
\xi^\bz(u,r,z,\bz) &=  \xi^\bz_0 (u,z,\bz) - \frac{1}{r}  \nabla^\bz \xi^u  + \mathcal{O}(r^{-2}) ,
\nn \\
\xi^u (u,r,z,\bz)&=  \xi^u_{0} (u,z,\bz) ,
\nn \\
\xi^r (u,r,z,\bz)&=- \frac{r}{2}  \nabla_A \xi^A_0  + \demi \nabla_A \nabla^A \xi^u + \mathcal{O}(r^{-1}) .
\end{align}
 $\nabla_A$ is the covariant derivative with respect to the unit $2$-sphere, giving  for example $\gamma_{z\bz} \nabla^z \xi^u = 2  \bp \xi^u$ and $\gamma_{z\bz}  \nabla_A \nabla^A \xi^u = 4  \pa_z \bp \xi^u$, etc....

 The subset of  the  $\Diff_4$  reparametrization symmetry  with   infinitesimal parameters   represented as   in \eqref{ghosts_without_EoM}     defines the leftover gauge  invariance of the gravitational action expressed   in the Bondi   gauge.  This  proves that it doesn't provide a full  gauge  fixing, as announced. 

Let us now take into account the falloff restrictions \eqref{s_falloffs}, which is to reduce further the 
coordinate  dependence  of the ghost zero modes.
 At leading order in the Bondi gauge, one can rewrite \eqref{sm_exact}, \eqref{smuzu_exact} and \eqref{sM_exact} of Appendix[\ref{Annexe_D}] as
\begin{align}
\label{smleading}
s \m &= \bp c^z + c^z \pa_z \m - \m \pa_z c^z + c^0 ( \pa_0 + \pa_\t) \m  + c^u  \pa_\t \m + \mathcal{O}(r^{-2}),
\\
\label{smuzuleading}
2s \mu^z_u &= s (\mt - \mo) = \pa_\t c^z - \pa_0 c^z +c^\bz (\pa_0 - \pa_\t )\m  - \frac{4}{t^2 \gamma_{z\bz}} \bp c^0
+ \mathcal{O}(r^{-2}),
\\
\label{sMleading}
2 s \mathcal{M} &= s(M+N)=\pa_\t c^\t +  \pa_0 c^0 
+ \mathcal{O}(r^{-2}).
\end{align}
One must observe   at this point  that  the $s$ transformations of the Beltrami fields are  computed 
in the coordinate system  $(\t,t,z,\bz)$ 
in Appendix[\ref{Annexe_D}].  Therefore,   one    must carefully  compute  the coordinate  derivations   on the fields  in \eqref{7.14}, consistently with  the   change of coordinates   $(\t,t,z,\bz) \to  (u,r,z,\bz  )$, that is 
\begin{align}
\t &\to \t = u + r 
\nn \\
x^0\equiv t &\to t = r  
\nn \\
\pa_\t &\to \pa_\t = \pa_u
 \nn \\
 \pa_0 &\to \pa_0 = \pa_r - \pa_u \ .
\end{align}
One   must also redefine the vector ghosts that are defined in Appendix[\ref{Annexe_D}] as 
\begin{align}
c^u &= c^\t - c^0 = -a \xi^0 + M \xi^\t - N \xi^0 - a \xi^\t = \mathcal{M} \xi^u ,
\nn \\
c^r &= c^0 = N \xi^0 + a \xi^\t = \mathcal{M} \xi^r + \frac{\mathcal{M}\mpm}{2} \xi^u ,
\nn \\
c^z &= \xi^z + \m \xi^\bz + \mu^z_u \xi^u ,
\nn \\
c^\bz &=  \xi^\bz + \mb \xi^z + \mu^\bz_u \xi^u.
\end{align}
 The  restrictions on the  ghosts  $\xi^\mu$ given by \eqref{xi_BMS_before_s_metric} and  the falloffs \eqref{power_expansion}  of the Beltrami fields can  then   be   safely combined to enforce  the falloff consistency conditions \eqref{s_falloffs} on \eqref{smleading}-\eqref{sMleading}.  One  gets
\begin{align}
\label{falloff_consistency}
0 &= (s \m)_0 = \bp \xi^z_0  \Rightarrow \xi^z_0 = \xi^z_0(z,u) ,
\nn \\
0 &= (s \mu^z_u)_0 = \pa_u \xi^z_0 \Rightarrow \xi^z_0 = \xi^z_0(z) ,
\nn \\
0 &= 2 (s \mathcal{M})_0 = \pa_u \xi^u + \xi^r_1 \Rightarrow \xi^u = \alpha(z,\bz) + \frac{u}{2} \nabla_A \xi^A_0, 
\end{align}
where $\alpha(z,\bz)$ is an arbitrary  given function of $(z,\bz)$ obtained by a trivial  quadrature  over the~$u$~coordinate.
Analogously, one gets   
  $\xi^\bz_0 = \xi^\bz_0(\bz)$. 
  
At this stage, one can simply check that all  remaining constraints associated to  \eqref{s_falloffs} are    satisfied 
   due to    \eqref{xi_BMS_before_s_metric} and \eqref{falloff_consistency}.    
It follows that  the  residual symmetry of  $\Diff_4$  in the Bondi gauge with the appropriate falloffs \eqref{fall-off_conditions}  is 
     carried  by the  restricted  four ghosts $\xi^\mu_{\rm BMS}$,  that   are  actually  well defined functionals of  a  ``fundamental   geometrical ghost representation" made of the   following anticommuting fields 
\begin{equation}
\label{extended_BMS_algebra}
 { \alpha(z,\bz) } \ , \hspace{0.5cm}  { \xi^z_0(z) } \ , \hspace{0.5cm}  { \xi^\bz_0(\bz) } .
\end{equation}
One can in fact  use the following notation
\begin{equation}
 {\xi_{\rm BMS} = \xi (\alpha,\xi^A_0,g_{\rm Bondi})} 
\end{equation}
with
\begin{align}
\label{final_BMS_ghosts}
\xi^u _{\rm BMS}&= \alpha(z,\bz) + \frac{u}{2}  \nabla_A \xi^A_0  ,
\nn \\
\xi^z_{\rm BMS} &= \xi^z_0(z)  - \frac{1}{r} \nabla^z \xi^u_{\rm BMS} + \mathcal{O}(r^{-2})  ,
\nn \\
\xi^\bz _{\rm BMS}&= \xi^\bz_0(\bz)  - \frac{1}{r} \nabla^\bz \xi^u_{\rm BMS} + \mathcal{O}(r^{-2})   ,
\nn \\
\xi^r_{\rm BMS} &= - \frac{r}{2}  \nabla_A \xi^A_0  + \demi \nabla_A \nabla^A \xi^u_{\rm BMS} + \mathcal{O}(r^{-1}) .
\end{align} 

The BMS beautiful result is  of course recovered by   expanding  in  spherical harmonics  the   arbitrary function    $\alpha(z,\bz)$, namely      the  non trivial  asymptotic  $d=4$ reparametrization invariance in the Bondi gauge includes not only the obvious  spatial and time translations through the lowest harmonics ($l=0,1$) but  the so-called {supertranslations} also arise  as the higher order harmonics $l,m$ for $l>1, m=-l,...,l$.
 
      $\xi^z_0$ and $\xi^\bz_0$ can be identified as solutions of a  two-dimensional conformal Killing equation  on the two-dimensional celestial sphere.  As noted elsewhere, it is well known that the set of  globally well defined solutions to this equation on $\mathcal{S}^2$ is six-dimensional, and the resulting algebra is  equivalent to that  of the $d=4$  Lorentz group. But if one allows local solutions with singularities, the set of solutions is enhanced to form the infinite-dimensional Virasoro  algebra which generates the  {superrotations}.  In this case, \eqref{extended_BMS_algebra} generate the  so-called  {extended BMS4 algebra}.

The  defining   horizontality conditions \eqref{topological_torsion_main}  actually  contains a built in representation of this algebra through the action of the BRST operator $s$ on the primary ghosts \eqref{extended_BMS_algebra}. In fact, rewriting \eqref{scz_exact} and \eqref{scu_exact} at leading order in the Bondi gauge leads to
\begin{align}
sc^z &= c^z \pa_z c^z + \mathcal{O}(r^{-1}),
\nn \\
sc^u &= s (c^\t - c^0) =  c^\bz \bp c^u + c^z \pa_z c^u -  \demi c^u \pa_\t c^0   - \demi c^u \pa_0 c^\t     + \mathcal{O}(r^{-2}) .
\end{align}
That is
\begin{align}
s\xi^z_0 &= \xi^z_0 \pa_z \xi^z_0 ,
\nn \\
s\xi^u &=   \xi^A_0 \pa_A \xi^u + \xi^u \pa_u { \xi^u } = \xi^A_0 \pa_A \xi^u + \demi \xi^u \nabla_A \xi^A_0 .
\end{align}

Using the same technique for $sc^\bz$ and decomposing $\xi^u$ as in \eqref{final_BMS_ghosts}, one gets the nilpotent BRST operator of the extended BMS4  algebra:
\begin{align}
\label{BRSTBMSoperator}
s \alpha &=\xi^A_0 \pa_A \alpha + \frac{\alpha}{2} \nabla_A \xi^A_0 
\nn \\
s\xi^z_0 &= \xi^z_0 \pa_z \xi^z_0
\nn \\
s\xi^\bz_0 &= \xi^\bz_0 \bp \xi^\bz_0.
\end{align}
Note that this algebra is realized at the level of the $\xi^\mu_{\rm BMS}$ given by \eqref{final_BMS_ghosts} when equipped with the modified BRST operator $\tilde s \equiv s - \delta^g_\xi$ where $\delta^g_\xi$ was introduced in \cite{BMS/CFT}. It measures the variation of the $\xi_{\rm BMS}$'s under a diffeomorphism due to their explicit dependence in the metric and acts as $\delta^g_\xi g_{\mu \nu} = \Lie_\xi g_{\mu \nu}$ on the metric.


 An important point   is understanding the action of a BMS transformation on the relevant  data at null infinity. In the Bondi language,  it amounts to compute the transformation laws 
of  the shear  tensor  and its derivatives   under the just built BRST--BMS operator. 
In terms of Beltrami fields, the shear tensor writes 
\begin{align}
  \label{shear_Beltrami}
C_{AB} &=
\gamma_{z\bz}
\pbM
{\mb}_{-1} & 0 
\\
0 & {\m}_{-1}
\peM.
\end{align}
This equation  is one of the many indications  that  the fields ${\m}_{-1}$ and ${\mb}_{-1}$ stand for  field representations of the two helicity states of the graviton.  More on that will appear in  a different publication.   The $s$ transformation    of the shear is   given by the term of order $\mathcal{O}(r^{-1})$ in \eqref{smleading}, that is 
\begin{align}
s{\m}_{-1} &= \bp \xi^z_{-1} + (\bp {\m}_{-1} ) \xi^\bz_0 + \xi^z_0 \pa_z {\m}_{-1} - {\m}_{-1} \pa_z \xi^z_0 + \xi^u \pa_u {\m}_{-1}  - \xi^r_1 {\m}_{-1},
\end{align}
which eventually gives 
\begin{align}
\label{s_shear}
s{\m}_{-1} &= \Big(  \xi^u \pa_u + \xi^A_0 \pa_A + \frac{3}{2} \bp \xi^\bz_0 - \demi \pa_z \xi^z_0 \Big){\m}_{-1} - \bp^2 \xi^u  
\end{align}
when the freedom on the determinant gauge condition ($\o'$ in \eqref{s_falloffs}) is used to fix the conformal factor $\exp \Phi$ such that $\gamma_{z\bz} \exp \Phi = 1 + \mathcal{O}(r^{-2}) $. The transformation $s{\mb}_{-1}$ is of course analogous to this one.

 The above computed values of the transformation of the Beltrami differential  $s{\m}_{-1}$ and  $s{\mb}_{-1}$  
 in \eqref{s_shear} and of the primary 
 ghosts   $\xi^z_0, \xi^\bz_0$   and $\alpha$ in \eqref{BRSTBMSoperator}
 coincide with the transformations  of the  BMS4   algebroid   as it is   computed  by Barnich in the Hamiltonian formalism \cite{Barnich_BRST}.

 What differs in our derivation is that it directly and unambiguously computes    the nilpotent BRST symmetry for  the BMS4 algebra from 
 the   geometrical  four-dimensional horizontality conditions \eqref{topological_torsion_main}, which express the nilpotent BRST symmetry for the 
 full
 $\Diff_4$, suitably constrained by the  four  Bondi gauge conditions and the falloff conditions  of the ten Beltrami fields.     
 
 Getting this  straight   four-dimensional derivation,  allowed   by the use of the  Beltrami parametrization, is an important result      of this work.  We   actually  expect that the Beltrami  formalism is   going   to be very helpful  for handling  the non trivial situation  of computing the BMS structure when local supersymmetry is involved,  through the horizontality conditions of  topological gravity with  further twists. This will appear in a separate publication.
 
Having clarified the BRST structure of  the BMS4 symmetry  allows us to pass to the next section devoted to  the  search  of non trivial solutions of the Wess and Zumino equations for  the BMS4 symmetry, that is for the search of  non trivial consistent   cocycles for the  $s$ symmetry, postulated as a fundamental symmetry for defining  relevant three-dimensional    quantum field theories  on the boundaries $\cal I^\pm$ of
  $\mathcal{M}_4$.

\section{Non Trivial Cocycles of the Nilpotent  BRST BMS4 Algebra}
\label{Section4}

The  BRST BMS4  symmetry is a truncation of the full $\Diff_4$ BRST symmetry with  restricted  reparametrization  ghosts. One might call the $3$ ghosts
 $\xi^z(z), \xi^\bz(\bz), \alpha(z,\bz)$ 
  its ``fundamental"  ghost  field representation, defined by   \eqref{BRSTBMSoperator}.

The $\Diff_4$ BRST symmetry has no non trivial cocycles as can be proven  by
showing  their equivalence with hypothetical non trivial ones for the local Lorentz symmetry. Indeed,   the latter cannot exist  for  $SO(1,3)$ (more generally the   Lorentz anomalies  can  only exist for $d=2$ modulo-$4$).
However,  the cocycle equations for the BRST BMS4 symmetry operator $s$ are less demanding than those  for the full
$SO(1,3) \times \Diff_4$ symmetry and nothing but a computation can confirm if they have or not non trivial solutions. 
It is  a mathematically legitimate question to look for such solutions.

 Moreover, a better  understanding  of the gravitational boundary effects  suggests building  three-dimensional  local
quantum field theory in $\cal  I^+$ governed by the nilpotent BRST BMS4 symmetry as constructed in the last section,  a bit analogously as one builds string theory as a bidimensional theory with the 
remaining symmetry of the $\Diff_2$ of the worldsheet in the conformal gauge, namely the modular invariance.
The physical motivation  for computing  the cohomology of the BRST operation $s$  at various ghosts numbers
is for checking if quantum anomalies may occur. Therefore, one must check 
if  a local   $3$-form cocycle with ghost number $1$  exists, generically denoted as 
$\Delta^1_3  (   {\m}_{-1},  {\mb}_{-1}, \xi^z (z), \xi^\bz(\bz), \alpha(z,\bz)) $.

$\Delta^1_3 $ must be such that a $2$-form    $ \Delta_2^2$ cocycle with  ghost number $2$  exists and is solution of
\bea\label{cbcb}
s  \Delta^1_3 + d \Delta_2^2 = 0,
\eea
and so on  until a  possibly $s$ invariant  $ \Delta_4^0$ cocycle that  is not  $s$ exact.

 $\Delta^1_3$ being a non trivial cocycle  means that  both $\Delta^1_3$ and $ \Delta_2^2$    are  defined modulo $s$  and $d$ exact terms. If such a cocycle exists, 
$\int_{\mathcal{I}^+} \Delta^1_3 $ may appear in the right hand side of the Lagrangian   BRST BMS4 Ward identity times a non zero coefficient ${\it{\bf  a}}$ whose value depends on the chosen QFT in  $\cal  I^+$,  without being possibly 
   absorbed in the effective $3$-dimensional action by     appropriate  changes in its renormalization procedure.  This would  thereby    invalidate    the chosen   $3$-dimensional local 
quantum field theory that one would consider in
 $\cal  I^+$ based on the principle of  $s$ invariance.

\subsection{A Bidimensional Motivation}

To gain intuition   on the  way to compute non trivial cocycles for the BRST BMS4 symmetry, one may firstly consider 
the nilpotent  ghost transformations    \eqref{BRSTBMSoperator} of its basic ghost
$\xi^z_0(z) $,  $\xi^\bz_0(\bz) $    and $\alpha (z,\bz) $. As such, these fields can be considered as genuine bidimensional objects, with no reference whatsoever  to the  third and fourth coordinates  $r$ and $u$.

One may indeed 
 consider    a   quite simple    bidimensional structure, based on  the holomorphic and antiholomorphic  ghosts 
 $\xi^z_0  $ and $\xi^\bz_0  $. Denoting   $m^z_\bz (z,\bz) \equiv {\m}_0$ and 
 $m^\bz_z (z,\bz) \equiv {\mb}_0$, one  can define  the generalized anticommuting $1$-forms on the celestial sphere
 $  \tilde M_0^z \equiv  dz  +   m^z_\bz( z,\bz) d\bz  + \xi^z_0  (z)$   and 
  $  \tilde M_0^\bz \equiv  d\bz  +   m^\bz_z( z,\bz) dz  + \xi^\bz_0  (\bz)$.  Part 
 of the intuition for doing so is  that $m^z_\bz( z,\bz)$ can be interpreted as  the Beltrami  differential component 
  ${\m}_{ 0}( z,\bz)$ on  the celestial sphere, which has been consistently chosen equal to zero in our construction  of the extended BMS4 symmetry. 
  
The action of $s$ acting on 
  all  field components of $\tilde M^A$  is defined as  the following horizontality conditions
  \begin{align} 
  \label{horel}
(s+d ) \tilde M_0^z     - \demi \{    \tilde M_0^z , \tilde M_0^z    \}  _z &=0 ,
 \nn\\
 (s+d ) \tilde M_0^\bz     - \demi \{    \tilde M_0^\bz , \tilde M_0^\bz    \}  _\bz &=0.
\end{align}
$ \demi \{  ,    \}_A$    is the Poisson bracket  along the direction $x^A$. By expansion in ghost number, one    recovers  $s\xi^A= \xi^A\pa_A \xi^A = \demi  \{\xi^A , \xi^A \}_A  $ and one has
  $sm^z_\bz = \{\xi^z_0 , m^z_\bz\}_z $ and  $sm^\bz_z = \{\xi^\bz_0 , m^\bz_z\}_\bz $, keeping in mind that these fields are taken equal to zero elsewhere  in the paper.  These transformation laws are indeed the same than the one for the zero order component of the metric, ${\m}_0$ and ${\mb}_0$, when $c^A \equiv \xi^A_0$.
  Therefore, 
 these equations remain consistent with  $s{\m}_0= s{\mb}_0=0$ when ${\m}_0={\mb}_0=0$.
So \eqref{horel} consistently defines  the  action of the BRST BMS4 operation  $s$ on $\xi^z_0$  and  $\xi^\bz_0$ as well as on
$ {\m}_0$ and  ${\mb}_0 $. 

The nilpotency property $s^2=0  $ trivially  holds true  as  a   graded   nilpotent differential operation due to the Jacobi identity of the Poisson bracket and to the chosen grading properties of all fields.   

One may tentatively introduce  an horizontality condition for the  $s$ transformation of $\alpha$, by introducing  the following $1$-form
 $\tilde   U \equiv du    +U_z (z,\bz) dz+U_\bz (z,\bz) d\bz   +   \alpha$ such that  
\bea \label{magicalU}
\tilde G  \equiv     (s+d )\tilde   U    -   \{   \tilde {M}_0^A , \tilde U        \} _A
  ^{\lambda} 
   \eea
    where $ \{\tilde M_0^A , \tilde U \}^\lambda_A \equiv \tilde{M}_0^A\pa_A  \tilde{  U} + \lambda   \tilde{ U}  \nabla _A \tilde{M}_0^A $ and 
     $\lambda$ is any given  real number. 
  Both components of the 1-form  $U$,   $ U_z (z,\bz) \equiv {\mu^u_z}_0$ and    $ U_\bz (z,\bz) \equiv {\mu^u_\bz}_0$,      are  defined on the celestial sphere with no $u$ dependence.  $U$ can be understood as a BRST antecedent   of
   $\alpha$, according to $U\to U+\alpha$.
 
 Quite remarkably, the   part of ghost number $2$ of the horizontality condition $\tilde{G}=G$  provides 
  $s\alpha= \xi^A_0 \pa_A \alpha +  \lambda \alpha \nabla _A \xi ^A_0$, 
    so  \eqref {magicalU} 
provides  the $4$-dimensional  formula  \eqref{BRSTBMSoperator}  for $\lambda=\demi$.
The part with ghost number $1$  of  $\tilde{G}=G$ defines the $s$ transformation of the BRST antecedent  $U$ of $\alpha$, as follows
\bea
sU= -d\alpha +  \{\xi^A, U   \}^\lambda _A   
+ \{M_0^A , \alpha  \}_A ^\lambda   .
 \eea
One can check   that  $s^2 =0$   on  $ \alpha$, 
$U_z$ and $U_\bz$  for all values  of $\lambda$.      In fact,  this property   
is directly implied by  the Bianchi identity 
\bea
 d G  =        \{     M^A_0 , G      \} _A  ^{\lambda} \eea
since by using $\tilde G=G$, one has   identically
  \bea  \tilde d G  = s^2    \tilde U +  \{    \tilde M^A_0 , G      \} _A  ^{\lambda}, \eea
denoting $\tilde{d} = s+d$. The terms of ghost number larger or equal to $2$ of this equation then implies $s^2\tilde U =0$.
 
The choice     $ \lambda =\demi$  fixes the conformal  weight of $\alpha$ on the celestial sphere. It must be understood that this value is determined  by  using  all ingredients of the $\Diff_4$ symmetry near 
 $\cal I^+$, while   the $\lambda$   dependent formula \eqref{magicalU} are genuinely bidimensional and covariantly well defined on the celestial sphere with $du$ considered as a $0$-form, see \cite{Baulieu_Grimm}.
 
  The strength of  \eqref{horel} and \eqref{magicalU} is its simplicity for defining the $s$ operation acting on   $\alpha$ and $\xi^A_0$,
  and identifying their BRST antecedent           $ {\mu^u_z}_0$,  $ {\mu^u_\bz}_0$,   $ {\mu^z_\bz}_0$
  and $ {\mu^\bz_z}_0$.
   However, we have not been able  to reexpress 
  the  action of the  BRST BMS4 symmetry  on  the  graviton fields  
${\m}_{{-1}}$ and ${\mb}_{{-1}}$ as stemming from  purely bidimensional  horizontality equations analogous to   \eqref{horel}.   
The
 difficulty originates from  the terms  $(\xi^u \pa_u + \frac{3}{2}\bp \xi^\bz_0 - \demi \pa_z \xi^z_0){\m}_{-1} \subset s{\m}_{-1}$   in the full four-dimensional formula  \eqref{s_shear}.    Finding this term would require to modify \eqref{horel} as
 \begin{equation}
 \tilde{d} \tilde{M}^z = \tilde{M}^z \pa_z \tilde{M}^z + \tilde{U} \pa_u \tilde{M}^z - \tilde{M}^z \pa_u \tilde{U}
 \end{equation}
 where the generalized $1$-forms are now defined as
 \begin{align}
 \tilde{M}^z &\equiv dz + \frac{1}{r} {\m}_{-1} d\bz + \xi^z_0 + \frac{1}{r} ( \xi^z_{-1} + {\m}_{-1} \xi^z_0  ),
 \nn \\
 \tilde{U} &\equiv du + U_z dz + U_\bz d\bz + \alpha + \frac{u}{2} \pa_A \xi^A_0.
 \end{align}
 The term with ghost number one, proportional to $d\bz$ and of order $\mathcal{O}(r^{-1})$ would then imply the correct $s$ transformation for ${\m}_{-1}$, but the terms with ghost number $2$ would break the transformation law $s\xi^A_0$ of the primary ghosts.
 
The above results indicate the deep four-dimensional origin of the BMS symmetry of $\cal I^+$, despite its apparently simple two and three-dimensional structure. 
As a matter of fact, the celestial holography  program \cite{Pasterski_celestial_holo} could encounter difficulties   due to the necessity of making it   consistent with the complete $\Diff_4$ invariance of the bulk.
  
However, the bidimensional approach  \eqref{horel}
 suggests the following method for computing 
non trivial BRST BMS4 cocycles involving     the graviton field ${\m}_{{-1}}$ and ${\mb}_{{-1}}$.
One first  observes  that,  in terms of  the not so relevant  $m^z_\bz  = {\m} _{0}   $ and  
$m_z^\bz=   {\mb} _{0}  $, one can define the following unified invariant cocycle for the 
$d+s$ operation
\begin{equation}\label{co1}
\tilde{\Delta}_{4 (0)} \equiv {du } ( \tilde M ^z_0  \pa_z   \tilde M ^z _0  \pa_z^2  \tilde M ^z_0   -z\leftrightarrow \bz ).
\end{equation}
One has indeed  
\bea \tilde d \tilde \Delta_{4 (0)}=0,   \quad    \tilde{\Delta}_{4 (0)} \neq  \tilde{d}\tilde   \Delta_{3 (0)}     \eea
since \eqref{horel} and the properties $\tilde M ^A_0 \tilde M ^A_0 =0$, 
 $\tilde \pa_A M ^A_0 \tilde \pa_A M ^A_0 =0$ imply
 $  \tilde d   (\tilde M ^A_0  \pa_A   \tilde M ^A _0  \pa_A^2  \tilde M ^A_0 )
=0$. 
The ghost expansion of \eqref {co1}  provides       four
consistency equations    $ s  \Delta^{g}_{ 4-g  (0) }+d  \Delta^{g+1}_{3-g (0)} =0$ 
 with ghost numbers $1,2,3$ and $4$ 
 within $\cal I^+$, such that  the     $ \Delta^{g}_{ 4-g  (0)}$'s are 
 defined modulo $d$ and $s$ exact terms. For example, $\Delta^1_{3(0)}$ decomposes as follows
\begin{equation}
\Delta^1_{3(0)} = 2 du \w dz \w d\bz \ (\pa_z   \xi ^z_0 \pa_z ^2  {\m }_0 + \pa_\bz  \xi ^\bz_0  \pa_\bz^2   {\mb }_0). 
\end{equation}

  The  three-dimensional $s$ consistency  equation  $s \int_{\cal I^+} \Delta^1_{3(0)}=0$ indicates that 
one may  possibly get  an anomaly for   $d=3$  quantum field theories    built 
 within  $\cal I^+ $ from the principle of  the  $s$ invariance,  having set    ${\m}_{ 0}( z,\bz)=0$ in the derivation of the   BMS4 symmetry.  On the  other hand, the $u$~independence of ${\m}   _{0}$ and   ${\mb}   _{0}$ makes the  finality of the above exercise a bit formal, but the latter suggests how to find non trivial cocycles  solution for the  obviously more  interesting case of finding  anomalies for    $d=3$  quantum field theories 
involving the asymptotic graviton fields ${\m}   _{-1}$ and   ${\mb}   _{-1}$.

\subsection{Computing the BMS4 Cocycles in $\mathcal{I}^+$}

As a matter of fact, what concretely matters is  the possibility of an anomaly occurring when   computing  the various entities relative to the production  or absorption of gravitons in the celestial sphere,
whose  fields are represented  by ${\m}   _{-1}$ and   ${\mb}   _{-1}$, 
 while respecting all the consequences of the BMS4 symmetry, in particular for all conservation laws, including of course the celebrated Bondi mass loss formula (to be reformulated in a separate paper  using the Beltrami formulation).

One is thus    looking for a series of $s$-cocycles 
$\Delta ^g_{4-g      }$, $g=1,2,3,4$, possibly generated by  \eqref{cbcb}, which satisfy
\bea
\label{WZC}
s\Delta ^g_{ 4-g    } +d \Delta ^{g+1}_{3-g    }  =0
\eea
and 
  are  a priori dependent    {\it  (i)}
on the three ghosts $\xi_0^A$ and/or  $\alpha$, 
to possibly carry  the ghost number
  $g\neq 0 $ 
  of  $\Delta ^g_{  4-g     }$,  
 and  {\it (ii)} 
on  ${\m}   _{-1}$  and  ${\mb}   _{-1}$.
One assumes that   $\Delta^1_{ 3   }$ is made of   exterior products of  
$dz$, $d\bz$ and $du$ 
in order that it is a well defined form on $\cal I^+$.

Solving \eqref{WZC} is of course non trivial for   ${\m}   _{-1}$  and ${\mb}   _{-1}$  dependent cocycles, 
since    the  BRST BMS4   transformation law   \eqref{s_shear}     of the former fields seems not to be    expressible  as  part of a  genuine  horizontality condition implying  simple descent equations.

 However, 
if  one manages to  compute
$\Delta ^1_{3 } \propto dudzd \bz $ 
with a linear ghost   dependence  in $\xi^A_0$ and/or   $\alpha$ to carry ghost number one,  and satisfying
\bea ds \Delta ^1_{3 } =  -sd  \Delta ^1_{3  } =0 , 
\eea
the algebraic Poincar\'e lemma implies the existence  
of a $2$-form  $\Delta ^{2}_{2 }$ such that
\bea\label {cocy1}
\label{trr}
s\Delta ^1_{3 } +  d\Delta ^{2}_{2 }  =0
\eea
and then using the  now obvious equation  $ ds \Delta ^{2} _{2 }=0$ and so on,   one   gets successively the existence of non trivial cocycles satisfying
\begin{align} \label{cocycocy}
s\Delta ^2 _{2 } + d \Delta ^{3}_ {1 }  &=0 , 
\nn \\
s\Delta ^3_{1    }+ d \Delta ^{4}_{0 }  &=0 , 
\nn \\ 
s\Delta ^4_{0  }   &= 0.
\end{align}
Therefore, everything boils down to guessing a possible form for $   \Delta^1 _{   3 } $.
Going back to the above            ${\m }_0 $ and    ${\mb }_0 $ dependent      $s$-cocycle $   \Delta^1 _{   3(0) } = 2du\w dz\w d\bz \ ({\m }_0 \pa_z^3   \xi ^z_0  + {\mb }_0 \pa_\bz^3  \xi ^\bz_0) $, one can observe that, not only its derivation  was suggestive,     but the way  the   conformal indices get contracted to  get a $3$-form  is quite unique. 
It is indeed expected that a possible anomaly
$ \Delta ^1_{3   }$ should be linear  in the graviton fields  as $ \Delta ^1_{3  (0)     }$.

 This makes it quite natural   to check if the following equation is valid:
\bea\label{solsol}
  sd \Delta^1 _{  3      } =0 \quad {\rm with} \quad 
  \Delta^1 _{  3      }   =  du\w dz\w d\bz    \ ({\m }_{-1} \pa_z^3   \xi ^z_{0}    +  {\mb }_{-1}  \pa_\bz^3 \xi ^\bz_0    ).
    \eea
    
 To do so, one  must  uses \eqref{BRSTBMSoperator},\eqref{s_shear}, and,  quite amazingly, one finds that  $ sd (\Delta^1 _{  3  }) =0$ and thus  \eqref{cocycocy} follows.

    
  Getting the explicit   expressions of the cocycles  $\Delta^g _{   4-g  }$ for $g>1$   is rather hard work in the absence of horizontality  equations. As a matter of fact, 
\eqref{trr} and \eqref{cocycocy}  are    satisfied for 
 \begin{align}\label{solsolsol } 
{\Delta^2_2 }  =& \ dz\w  d\bz \  \xi^u ( \mu \pa^3 \xi + \bar{\mu} \bar{\pa}^3 \bar{\xi}  )    - du \w d\bz \  \big( \xi ( \mu \pa^3 \xi + \bar{\mu} \bar{\pa}^3 \bar{\xi}  ) + \pa \xi^u \bar{\pa}^3 \bar{\xi} \big) + du\w  dz \ \big( c.c. \big),
\nn \\
{\Delta^3_1 } =&\   - du \  \big(\xi \bar{\xi} ( \mu \pa^3 \xi + \bar{\mu} \bar{\pa}^3 \bar{\xi} ) + \xi (\pa^3 \xi) (\bar{\pa} \xi^u) - \bar{\xi} (\bar{\pa}^3  \bar{\xi}) (\pa \xi^u )  \big)
\nn \\
&\ - d\bz \ \xi^u \big( \xi (\mu \pa^3 \xi + \bar{\mu} \bar{\pa}^3 \bar{\xi} ) + \pa \xi^u \bar{\pa}^3 \bar{\xi} \big)  +dz 
\ \xi^u (c.c.)   )  , 
\nn \\
{\Delta^4_0}  =&\   \xi \pa^3 \xi (\bar{\xi} \xi^u \mu  + \xi^u \bar{\pa} \xi^u ) - {\bar{\xi}} \  {\bar{\pa}^3} \bar{\xi}  (  \xi \xi^u \bar{\mu}  + \xi^u \pa \xi^u )
\end{align}
with the simplified notations 
\begin{align}
 \mu &={\m}_{-1} , \quad \quad    \bar{\mu}={\mb}_{-1} ,   \quad \quad  \pa=\pa_z, \quad \quad   \bar{\pa}=\bp, \quad \quad  \xi=\xi^z_0, \quad \quad   \bar{\xi}=\xi^\bz_0.
\end{align}
The   last cocycle satisfies    $s \Delta_{0}^4 = 0$.  
The complicated expressions of  the lower descendant cocycles  $\Delta^2_{2}$, $\Delta^3_{1}$   and  $\Delta^4_0   $ contrast with the simplicity of
their generating  ghost number   one  up ascendant
$\Delta^1_{3}  
=du\w dz\w d\bz \ (   \mu \pa^3 \xi + \bar{\mu} \bar{\pa}^3 \bar{\mu}  )$,   as    (correctly) postulated in \eqref{solsol}. As a matter of fact, the simplicity of  the quadratic field dependence of $\Delta^1_{3}$   makes it rather obvious
to check 
that it  is neither $d$ or $s$ exact.  Hence,  one has the same property for
the other  cocycles $\Delta^g_{4-g} $.  So,  
the existence of $\Delta^1_{3} $ is a clear signal that one may have an anomaly in the context of  
three dimensional   
 Lagrangian quantum field theories in $\cal I^+$, relying on the principles of the BRST BMS4  invariance. 

In practice,
 ${\it{\bf  a}} {\it \cal  }\int _{\cal I^+ } {\Delta^1_3} (\xi, \bar{\xi}, \mu, \bar{\mu} )$   with  
 ${\it{\bf  a}} \neq 0$
 is   an    obstruction for imposing the Ward identity of the BRST BMS4 symmetry of the chosen 
 quantum field theories  in 
 $\cal I^+$.
 The study of this possibly broken Ward identity permits one to identify which correlation functions must be investigated  to compute the  value of 
 the model dependent anomaly coefficient ${\it{\bf  a}}$.
 Physically, having $ {\it{\bf  a}}\neq0$ may 
 complicate the construction  of theories describing graviton  absorptions and creations  in the null  boundaries $\cal I^\pm$ of asymptotically flat manifolds.


Alternatively, if one uses  a Hamiltonian formalism,  the coefficient 
${\it{\bf  a}}$ is to be interpreted as a central charge.  The  
 breaking   of the  Hamiltonian
BRST BMS4  Ward identity implies  that    of the nilpotency
of  its   BRST charge $Q$, as shown  e.g. by Kato and Ogawa \cite{Kato} in the context of the Polyakov covariant bosonic string theory.  One  actually expects   
$Q^2 =  {\it{\bf  a}}   \int_{\mathcal{S}^2} \Delta^2_{2 }$  in operational form.
It~is~noteworthy  that  the  $\Delta^g_{4-g}$'s  for~$g \geq 2$~correspond to   the   BMS4   algebroid cocycles     that  Barnich computed in \cite{Barnich_BRST}.   However,  the  Hamiltonian approach   leaves  undetermined  the generating   Lagrangian top cocycle $\Delta^1_{3} $.


%

\newpage

\appendix
\addcontentsline{toc}{part}{Appendices}
\part*{Appendices}
   
\section{Computation of the $d=4$ Beltrami Spin Connection}
\label{Annexe_A}

This appendix exposes the full form of the spin connection $\o(e)$ for $d=4$ Beltrami gravity as the solution of $24$ linear equations. They are the result of the vanishing torsion conditions $T^\t =T^0 = T^z = T^\bz = 0$ where $T^a = de^a + \o^a_{\ b} \w e^b$ and will be developed here. One works with the Beltrami parametrized vierbein \eqref{Bele}, namely
\bea
\label{basis}
\pbM
\Eo \cr \Eob \cr e^0 \cr e^\t
\peM
\equiv
\pbM
 1 & \m &  \mo & \mt  \\ \mb & 1  & \mbo & \mbt 
\\ 0 & 0 & N & a
\\
0 & 0 & \bar{a} & M
\peM
\pbM
dz \cr d\bz \cr dt \cr d\t
\peM
\eea
where $e^z = \exp \vp \ \mathcal{E}^z$, $e^\bz = \exp \vpb \ \mathcal{E}^\bz$ and $a\equiv \mu^0_\t$, $\bar{a} \equiv \mu^\t_0$.

The vanishing torsion equations are first solved in the general case $\vp \neq \vpb$ and $a \neq \bar{a}$, but this Lorentz gauge fixing will be enforced at the end of the computations. The following equations are slightly different from the one derived in \cite{leafofleaf} as one uses the Minkowski metric \eqref{Mink_sphere} $ds^2 = -d\t^2 + dt^2 + t^2 \gamma_{z\bz} dz d\bz$ in coordinates $(\t,t,z,\bz)$.


There are $12$ equations that come from the $2$-form decomposition of $T^\t = T^0 = 0$. This decomposition will be done on the basis $(\mathcal{E}^z,\mathcal{E}^\bz,dt,d\t)$. In this basis, the exterior derivative takes the form 
\begin{equation}
\label{exterior_derivative} 
d =   d\t  \pa_\t+    dt\pa_0  + dz\pa_z  + d \bz \bp  = d\t {\cal D}_\t+  dt {\cal D}_0  + \Eo \Dz  +  \Eob    \Dbz
\end{equation}
and the spin connection decomposes as  
\begin{equation}
\o^{ab} = d\t \o^{ab}_\t+  dt \o^{ab}_0  + \Eo \o^{ab}_Z  +  \Eob   \o^{ab} _{\bar Z}.
\end{equation}
The  expressions  of the curly derivatives $\mathcal{D}$ are recovered by using \eqref{basis}, which gives 
\begin{align}
\label{derD}
\pbM  \Dz \\ \Dbz   \peM &\equiv \frac{1}{\mmb} \pbM  
\pa_z - \mb \bp
\\
\bp - \m \pa_z
\peM,
\\
\pbM  \Do \\ \Dt   \peM &\equiv  \pbM  \pa_0 \\ \pa_\t  \peM - \frac{1}{\mmb}  
\pbM
(\mo - \m \mbo )\pa_z +(\mbo - \mb \mo )\bp
\\
(\mt - \m \mbt )\pa_z +(\mbt - \mb \mt)\bp
\peM.
\end{align}

The $12$ equations coming from $T^\t=T^0=0$ are then pretty easy to derive, this goes as 
\begin{align}
\label{Tt}
 T^\t &\equiv     de^\t     + \demi t^2 \gamma_{z\bz} \o^{\t z}   \w e^\bz + \demi t^2 \gamma_{z\bz} \o^{\t \bz}   \w e^z +  \o^{\t 0}   \w e^0  
 \nn \\
 &=
  (\Eo \Dz  M +\Eob \Dbz M  + \Do M dt )      \w d\t  
  + (\Eo \Dz  \bar{a} +\Eob\Dbz \bar{a}  +  \Dt \bar{a} d\t )      \w dt
   \nn \\
& \ \ \  -     \demi t^2 \gamma_{z\bz}  {\exp \vpb } \   \Eob  \w   \o^{\t z }
-      \demi t^2 \gamma_{z\bz}  {\exp \vp } \   \Eo  \w   \o^{\t \bz }
+
   \o^{\t0} \w
  ( N dt + a d\t)
    \nn\\ 
 &= {  \Eo  \w  \Eob} \ t^2 \gamma_{z\bz}
(    
  \demi  \exp \vpb\ \o^{\t z}_Z - \demi  \exp \vp \ \o_{\bZ}^{\t \bz} ) +dt \w d\t ( \Do   M   -   \Dt   \bar{a}  - N \o^{\t 0}_\t   +   a  \o^{\t0}_0  )
     \cr 
& \ \  \
+\Eo\w d\t  (     \Dz M   - 
\demi t^2 \gamma_{z\bz}
{\exp \vp} \  \o^{\t \bz}_\t  
 +\o^{\t0}_Z a   )
+\Eob \w d\t  (    \Dbz M   - \demi t^2 \gamma_{z\bz} {\exp \vpb} \  \o^{\t z}_\t  
 +\o^{\t0}_{\bZ}  a      )
\nn \\
&\ \ \ 
+\Eo\w dt  (     N \o^{  \t 0}_Z  -   \demi t^2 \gamma_{z\bz} {\exp \vp} \ \o^{\t\bz }_0   + \Dz \bar{a}  )
+\Eob\w dt  (     N \o^{  \t 0}_{\bZ}   -  \frac{1}{2}t^2 \gamma_{z\bz} {\exp \vpb} \ \o^{\t z }_0    +\Dbz \bar{a}   ),
 \end{align}
 which gives the first $6$ equations. The others come from the same kind of decomposition for $T^0$, that is
 \begin{align}
 \label{T0}
 T^0 &\equiv     de^0 - \demi t^2 \gamma_{z\bz}  \o^{ z 0} \w e^\bz  + \demi  t^2 \gamma_{z\bz} \o^{0\bz} \w e^z -\o^{0\t} \w e ^\t
 \nn\\ 
 &=   
 (\Eo \Dz  N +\Eob\Dbz N  +   \Dt   N d\t )      \w dt  
 +(\Eo \Dz  a +\Eob\Dbz a  +  \Do   a dt )      \w d\t 
\nn \\
& \ \ \  +  t^2 \gamma_{z\bz} \frac  {\exp \vpb } {  2}   \Eob  \w   \o^{z0}   - t^2 \gamma_{z\bz} \frac{\exp \vp } {2} \Eo \w  \o^{0 \bz}    -\o^{0\t}   \w  (\bar{a} dt +M d\t)
      \nn\\
 &= \demi \ t^2 \gamma_{z\bz} {    \Eo  \w  \Eob} 
(    \exp \vpb \ \o^{ z0}_Z +\exp \vp \ \o_{\bZ}^{0 \bz}) +d\t \w dt ( \Dt   N
-  \Do a  - M \o^{\t 0}_0    - \bar{a} \o^{0 \t }_\t   )
     \nn \\
& \ \ \  
+\Eo\w dt  (     \Dz N   - t^2 \gamma_{z\bz} \frac{\exp \vp}{2} \o^{ 0\bz}_0    -\bar{a} \o^{0\t}_Z    )
+\Eob \w dt  (     \Dbz N   + t^2 \gamma_{z\bz} \frac{\exp \vpb}{2} \o^{z0}_0     -\bar{a} \o^{0\t}_{\bZ}  )
\nn \\
&\ \ \ 
+\Eo \w d\t  ( -    M \o^{ 0 \t }_Z  - t^2 \gamma_{z\bz} \frac{\exp \vp}{2} \o^{0\bz}_\t      +\Dz a   )
+\Eob \w d\t  ( -   M \o^{ 0 \t }_{\bZ}  +t^2 \gamma_{z\bz}  \frac{\exp \vpb}{2} \o^{z0}_\t    +\Dbz a 
   ).  
\end{align}
 
To find the $12$ remaining equations, one must do the same $2$-forms expansion for $T^z$ and $T^\bz$. This is slightly more involved than doing it for $T^\t$ and $T^0$ because some relations are needed to relate the components of forms expressed either on the basis of $1$-forms $(dz,d\bz,dt,d\t)$ or on the basis $(\Eo, \Eob,dt,d\t)$. This is done by rewriting all products of forms involving $dz$ and/or $d\bz$ in terms of the elements of the basis $(\Eo, \Eob,dt,d\t)$.  By using (\ref{basis}), it yields to
\bea
\pbM dz \\ d\bz \peM = \frac{1}{\mmb} \pbM  1 & -\m \\ -\mb & 1  \peM
\Big  [  
\pbM \Eo  \\ \Eob \peM 
 -\Mo
  \pbM dt  \\ d\t \peM 
  \Big ]. 
\eea
Then,  $T^z$ and $T^z$ read
 \begin{align}
 T^z &\equiv    de^z - \demi t^2 \gamma_{z\bz} \o^{\bz z}  \w e^z   +   \o^{z0} \w e^0 -\o^{z \t}  \w e^\t \nn 
 \\
 &= 
    \exp \vp \Big(
    ( d \vp   - \demi t^2 \gamma_{z\bz} \o^{\bz z}  ) \w \Eo   +d\Eo \Big) +             \o^{z0} \w( Ndt+ a d\t)
      -          \o^{z\t}     \w (\bar{a} dt +M d\t ) 
     \nn\\
     &= 
    \exp\vp
    \Big (  
    ( d \vp     - \demi t^2 \gamma_{z\bz} \o^{\bz z}  )\w\Eo 
     + \exp (-\vp)   \    \o^{z0} \w (Ndt  +a d\t)
       - \exp (-\vp)  \   \o^{z\t} \w    ( M   d\t      +\bar{a} dt) 
    \nn \\ &\ \ \ 
    + dz\w d\bz  \   \pa_z \m 
    +
     dt\w d\bz  \   (\pa_0 \m - \bp \mo)
    -dt\w  dz   \   \pa_z \mo    
     +    
    d\t \w d\bz  \ (  \pa_\t \m   -\bp \mu^z_\t)
      \nn \\
      &\ \ \ -d\t\w  dz   \   \pa_z \mu^z_\t
   +
      dt\w   d  \t
       \   (\pa_0 \mu^z_\t- \pa_\t \mo)  
       \Big),
     \end{align}
     \begin{align}
 T^\bz &\equiv    de^\bz + \demi t^2 \gamma_{z\bz} \o^{\bz z}  \w e^\bz   -   \o^{0\bz} \w e^0 -\o^{\bz\t} \w  e^\t \nn
 \\
 &= 
    \exp\vpb
    \Big (  
    ( d \vpb     + \demi t^2 \gamma_{z\bz} \o^{\bz z}  )\w\Eob   +d\Eob \Big) -     \o^{0\bz} \w( Ndt+a d\t)
      -      \o^{ \bz \t}     \w (\bar{a} dt +M d\t)    \Big )
     \nn\\
     &= 
    \exp\vpb
    \Big (  
    ( d \vpb     + \demi t^2 \gamma_{z\bz} \o^{\bz z}  )\w\Eob 
     - \exp (-\vpb)  \     \o^{0\bz} \w (Ndt  +a d\t)
       - \exp (-\vpb)  \   \o^{\bz\t} \w    (\bar{a} dt+  M   d\t      ) 
    \nn \\ &\ \ \  
    + d\bz\w dz  \   \bp \mb
    +
     dt\w dz  \   (\pa_0 \mb- \pa_z \mo)
    -dt\w  d\bz   \   \bp \mob    
     +    
    d\t \w dz  \ (  \pa_\t \mb   -\pa_z \mu^\bz_\t)
      \nn \\
      &\ \ \ -d\t\w  d\bz   \   \bp \mu^\bz_\t
   +
      dt\w   d  \t
       \   (\pa_0 \mu^\bz_\t- \pa_\t \mob)  
       \Big).
     \end{align}
The condition $T^z=0$ is then
      \begin{align} 
       \label{Tz}
     0  &=  \Eo\w\Eob
       (-\Dbz \vp + \demi t^2 \gamma_{z\bz} \o^{\bz z}_{\bZ}   +\frac{\pa_z\m}{\mmb} )
   \nn\\     &\ \ \ +
       \Eo\w d t
      \Big  (-\Do \vp + \demi t^2 \gamma_{z\bz} \o^{\bz z}_0    + \exp-\vp  ( N \o_Z^{z0} - \bar{a}  \o_Z^{z \t}   )      
      \nn \\
      &\ \ \ \hspace{1.55cm}+\frac{1}{\mmb}  
       (  \pa_z \mo
       -
       \mob \pa_z \m
       +\mb (\pa_0 \m- \bp \mo)
       )\Big)
       \nn\\
         &\ \ \ +  \Eob \w  d t  
         \Big(    \exp-\vp  ( N \o^{z0}_{\bZ}  -  \bar{a} \o^{z \t}_{\bZ} )    
         +\frac{1}{\mmb} (  - \pa_0 \m+\bp \mo +\mo \pa_z \m
-\m\pa_z\mo)
\Big)
       \nn\\
            &\ \ \ +
      \Eo \w  d\t
       \Big( 
       -   \Dt 
       \vp  + \demi t^2 \gamma_{z\bz} \o^{\bz z}_\t    +  \exp-\vp   (     -  M  \o^{z \t} _Z
       + a \o^{z0}_Z    )
       \nn \\
       &\ \ \ \hspace{1.65cm}+
       \frac{1}{\mmb}
       (  \pa_z \mt
       - \mbt \pa_z\m
       +  \mb  (\partial_{\t} \m- \bp \mu^z_\t) ) 
       \Big)
       \nn\\
           &\ \ \ +
            \Eob\w  d\t 
           \Big(\exp-\vp (
            a \o^{z0}_{\bZ}  -  M \o^{z \t}_{\bZ}
           )
           +     \frac{1}{\mmb}      (
            -   \pa_\t \m +\bp \mu^z_\t 
                + \mt  \pa_z \m 
           -\m \pa_z \mt)
       \Big)
\nn\\
           &\ \ \ + d t \w d\t
           \Big(
           \exp-\vp
           ( a  \o^{z0}_0   -  N  \o^{z0}_\t    - M \o^{z\t }_0
             +\bar{a} \o^{z\t }_\t  )
             +   \pa_0 \mu ^z_\t  -  \pa_\t \mu^z_0
              \nn\\  &\ \ \  \hspace{1.55cm} 
             +
              \frac{1}{\mmb} 
        \big[ \pa_z\m(\mo\mu^\bz_\t - \mu^z_\t   \mu^\bz_0)
        -( \pa_0\m-\bp\mo) (\mu^\bz_\t  -\mb \mu^z_\t )
        +\pa_z\mo(\mu^z_\t -\m\mu^\bz_\t )  
        \nn\\&\ \ \ 
        \hspace{3.56cm}
        +(\pa_\t \m-\bp\mu^z_\t)(\mbo-\mb \mo)
        -
        \pa_z\mu^z_\t (\mo-\m\mbo) \big]
            \Big)
\end{align}
from which $6$ other linear equations are deduced. 
  
Analogously, the condition $T^\bz=0$ writes 
 \begin{align}
 \label{Tbz}
    0  &= 
       \Eob\w\Eo
       \Big(-\Dz \vpb - \demi t^2 \gamma_{z\bz} \o^{\bz z}_Z   +\frac{ \bp\mb}{\mmb} \Big)
   \nn\\     &\ \ \ +
       \Eob\w d t
      \Big( -\mathcal{D}_{0} \vpb - \demi t^2 \gamma_{z\bz} \o^{\bz z}_0    -\exp-\vpb ( N \o_{\bZ}^{0\bz} +\bar{a}  \o_{\bZ}^{\bz \t} )      
      \nn \\
      &\ \ \ \hspace{1.55cm} +\frac{1}{ \mmb}  
       (  \bp\mob
       -
       \mo \bp\mb
       +\m (\pa_0 \mb- \pa_z \mob)
       )   \Big)
       \nn\\
         &\ \ \ +   \Eo \w  d t  
         \Big(  - \exp-\vpb  ( N \o^{0\bz}_Z  + \bar{a} \o^{\bz \t}_Z )    
         +\frac{1}{\mmb}(  - \pa_0 \mb+\pa_z \mob +\mob\bp\mb
-\mb\bp\mob)
\Big)
       \nn\\
            &\ \ \ +
       \Eob\w d\t
       \Big( 
       -  \mathcal{D}_{\t} \vpb -  \demi t^2 \gamma_{z\bz} \o^{\bz z}_\t    -  \exp-\vpb   (       M  \o^{\bz \t} _{\bZ}
       + a \o^{0\bz}_{\bZ}    )
       \nn \\
       &\ \ \ \hspace{1.64cm}+
       \frac{1} {\mmb}
       (   \bp\mbt
       - \mt \bp \mb
       +  \m  (\pa_\t \mb- \pa_z \mu^\bz_\t)
         ) 
       \Big)
       \nn\\
           &\ \ \ +
            \Eo\w d\t 
           \Big( - \exp-\vpb  (
           a \o^{0\bz}_Z  +  M \o^{\bz \t}_Z
           )
           +     \frac{1} {\mmb}      (
            -   \pa_\t \mb +\pa_z \mu^\bz_\t 
                + \mbt  \bp\mb 
           -\mb\bp\mbt) 
       \Big)
\nn\\
           &\ \ \ + d t\w d\t
           \Big(
          - \exp-\vpb
           (  a  \o^{0 \bz}_0   - N  \o^{0\bz}_\t  
             + M \o^{\bz \t}_0
             -\bar{a} \o^{\bz \t}_\t  ) +\pa_0\mu^\bz_\t - \pa_\t\mu^\bz_0 
              \nn \\  &\ \ \  \hspace{1.55cm}   
             +
             \frac{1}{\mmb} 
        \big[ -\pa_\bz \mu^\bz_0(\mb\mu^z_\t-\mu^\bz_\t) + (\pa_z\mu^\bz_\t-\pa_\t\mb)(\m\mu^\bz_0-\mu^z_0)   - \pa_\bz\mb(\mu^z_0\mu^\bz_\t-\mu^\bz_0\mu^z_\t)  \nn
        \\ &\ \ \ \hspace{3.56cm}
         - (\pa_z\mu^\bz_0-\pa_0\mb)(\m\mu^\bz_\t-\mu^z_\t) + \pa_\bz\mu^\bz_\t(\mb\mu^z_0-\mu^\bz_0)
          \big]  \Big)
\end{align}
from which one deduces the last $6$ equations that are necessary to compute $\o(e)$.

The full system of $24$ linear independent equations, made of (\ref{Tt}), (\ref{T0}), (\ref{Tz}) and (\ref{Tbz}), can be solved numerically for $\bar{a}=-a$ and $\vp=\vpb=\frac{\Phi}{2}$. One then obtains the expression of the $24$ components of the spin connection $\o(e)$ when the vierbein is expressed in its Beltrami form:
\newgeometry{top=1cm,bottom=2cm,left= 1cm,right=1cm}
\begin{align}
\label{Beltrami_sc}
\o^{\bz z}_0 &= \frac{1}{t^2 \gamma_{z\bz} } (\mathbb{D}_\bz \mu^\bz_0-\mathbb{D}_z \mu^z_0)
\nn \\
\o^{\bz z}_Z &= -\frac{1   }{t^2\gamma_{z\bz}} \mathbb{D}_z \Phi
 \nn \\
\o^{\bz z}_{\bZ}  &= \frac{1  }{t^2\gamma_{z\bz}}  \mathbb{D}_\bz \Phi 
  \nn \\ 
\o^{z 0}_0  &= \frac{-\exp(-\frac{\Phi}{2} )}{ t^2 \gamma_{z\bz} (a^2 +M N)} \left( a^2  \mathcal{D}_\bz N+ a^2 \mathcal{D}_\bz M+ a N \mathcal{D}_\bz a -a M \mathcal{D}_\bz a + 2 MN \mathcal{D}_\bz N \right) - \frac{a \exp(\frac{\Phi}{2})}{2(a^2  +MN )} \left(  \mathbb{D}_0 \mu^z_\t - \mathbb{D}_\t \mu^z_0  \right) 
  \nn \\
 \o^{0 \bz}_0  &= \frac{- \exp(-\frac{\Phi}{2}) }{t^2 \gamma_{z\bz} \left(a^2 + M N\right)} \left( -a^2 \mathcal{D}_z N- a^2 \mathcal{D}_z M  -a N \mathcal{D}_z a +a M \mathcal{D}_z a -2 MN\mathcal{D}_z N \right) + \frac{a\exp(\frac{\Phi}{2}) }{ 2 \left(a^2 + M N\right)} \left(  \mathbb{D}_0 \mu^\bz_\t  - \mathbb{D}_\t \mu^\bz_0  \right) 
   \nn \\
 \o^{\bz z}_\t   &= \frac{1}{t^2 \gamma_{z\bz}} ( \mathbb{D}_\bz \mu^\bz_\t  -\mathbb{D}_z \mu^z_\t)
    \nn \\
 \o^{z 0}_\t   &= \frac{\exp(-\frac{\Phi}{2} )}{ t^2 \gamma_{z\bz} (a^2+M N)} \left(-2 a^2  \mathcal{D}_\bz a - aM \mathcal{D}_\bz N + a M \mathcal{D}_\bz M - MN \mathcal{D}_\bz a -M^2 \mathcal{D}_\bz a \right)  +\frac{M\exp(\frac{\Phi}{2} )}{2  (a^2+M N)}  \left( \mathbb{D}_0 \mu^z_\t  -\mathbb{D}_\t \mu^z_0   \right) 
     \nn \\
\o^{0 \bz}_\t    &= \frac{2  \exp(-\frac{\Phi}{2})}{t^2 \gamma_{z\bz}} \mathcal{D}_z a + \frac{ M \exp( -\frac{\Phi}{2} ) }{t^2 \gamma_{z\bz} \left(a^2 + M N\right)} \left( a \mathcal{D}_z N  -a  \mathcal{D}_z M-  N \mathcal{D}_z a +  M \mathcal{D}_z a \right) + \frac{ M \exp(\frac{\Phi}{2} ) }{ 2 \left(a^2 +M N\right)} \left(  \mathbb{D}_\t \mu^\bz_0  - \mathbb{D}_0 \mu^\bz_\t  \right)
      \nn \\
 \o^{z 0}_{\bZ}     &= \frac{\exp(\frac{\Phi}{2})}{a^2+M N} (a \mathbb{D}_\t \m + M \mathbb{D}_0 \m )
       \nn \\ 
 \o^{0 \bz}_Z    &= \frac{-\exp(\frac{\Phi}{2}) }{a^2+M N} (a \mathbb{D}_\t \mb + M \mathbb{D}_0 \mb )
       \nn \\
 \o^{z 0}_Z      &= \frac{\exp(\frac{\Phi}{2}) }{2 (a^2+M N)} \Big( -a \big( \mathbb{D}_\bz \mu^\bz_\t+ \mathbb{D}_z \mu^z_\t  - \mathcal{D}_\t \Phi \big) + M \big( \mathcal{D}_0 \Phi  -\mathbb{D}_\bz \mu^\bz_0 -\mathbb{D}_z \mu^z_0 \big)\Big)
        \nn \\ 
  \o^{0 \bz}_{\bZ}     &= \frac{-\exp(\frac{\Phi}{2}) }{2 (a^2+M N)} \Big( -a \big( \mathbb{D}_\bz \mu^\bz_\t  + \mathbb{D}_z \mu^z_\t - \mathcal{D}_\t \Phi \big) + M \big( \mathcal{D}_0 \Phi -\mathbb{D}_\bz \mu^\bz_0 -\mathbb{D}_z \mu^z_0 \big) \Big)
        \nn \\ 
 \o^{z \t}_Z   &=  \frac{\exp(\frac{\Phi}{2}) }{2 (a^2+M N)}  \Big( a \big( \mathcal{D}_0 \Phi - \mathbb{D}_\bz \mu^\bz_0- \mathbb{D}_z \mu^z_0 \big) + N \big( \mathbb{D}_\bz \mu^\bz_\t   +\mathbb{D}_z \mu^z_\t  -\mathcal{D}_\t \Phi  \big) \Big)
        \nn \\ 
 \o^{\bz \t}_{\bZ}     &= \frac{\exp(\frac{\Phi}{2})}{2 (a^2+M N)}  \Big(a \big( \mathcal{D}_0 \Phi - \mathbb{D}_\bz \mu^\bz_0- \mathbb{D}_z \mu^z_0 \big) + N \big( \mathbb{D}_\bz \mu^\bz_\t   +\mathbb{D}_z \mu^z_\t -\mathcal{D}_\t \Phi \big) \Big)
        \nn \\ 
  \o^{z \t}_{\bZ}   &= \frac{\exp(\frac{\Phi}{2} ) }{a^2+M N}(a \mathbb{D}_0 \m-N\mathbb{D}_\t \m )
        \nn \\
  \o^{\t z}_\t    &= \frac{-\exp(-\frac{\Phi}{2})}{ t^2\gamma_{z\bz} (a^2+M N)}   \left(- a^2 \mathcal{D}_\bz N- a^2  \mathcal{D}_\bz M+ a N \mathcal{D}_\bz a - a M\mathcal{D}_\bz a -2MN \mathcal{D}_\bz M \right)
- \frac{a\exp(\frac{\Phi}{2} )}{2  (a^2+M N)} \left( \mathbb{D}_0 \mu^z_\t - \mathbb{D}_\t \mu^z_0  \right) 
         \nn \\
  \o^{\t \bz}_\t      &= \frac{- \exp(-\frac{\Phi}{2} )}{t^2 \gamma_{z\bz}\left(a^2 + M N\right)}  \left( -a^2  \mathcal{D}_z N - a^2  \mathcal{D}_z M + a N \mathcal{D}_z a - a M \mathcal{D}_z a -2MN  \mathcal{D}_z M \right) - \frac{a \exp(\frac{\Phi}{2})}{ 2\left(a^2 + M N\right)}  \left(  \mathbb{D}_0 \mu^\bz_\t  - \mathbb{D}_\t \mu^\bz_0   \right)
  \nn \\
   \o^{\bz \t}_Z    &= \frac{\exp(\frac{\Phi}{2}) }{a^2+M N} (a \mathbb{D}_0 \mb- N\mathbb{D}_\t \mb )
          \nn \\
    \o^{\t \bz}_0      &= \frac{-2  \exp(-\frac{\Phi}{2}  )}{t^2 \gamma_{z\bz}}  \mathcal{D}_z a + \frac{N \exp(-\frac{\Phi}{2}  ) }{t^2 \gamma_{z\bz} \left(a^2 + M N\right)}  \left( a  \mathcal{D}_z N -a  \mathcal{D}_z M- N  \mathcal{D}_z a + M  \mathcal{D}_z a \right)
   + \frac{N \exp(\frac{\Phi}{2} ) }{ 2\left(a^2 + M N\right)} \left(   \mathbb{D}_\t \mu^\bz_0   -\mathbb{D}_0 \mu^\bz_\t   \right)
           \nn \\
    \o^{\t z}_0    &= \frac{-\exp(-\frac{\Phi}{2} ) }{ t^2 \gamma_{z\bz}(a^2+M N)}  \left(2 a^2  \mathcal{D}_\bz a- a N\mathcal{D}_\bz N + aN \mathcal{D}_\bz M +N^2 \mathcal{D}_\bz a + MN\mathcal{D}_\bz a \right) - \frac{N \exp(\frac{\Phi}{2} ) }{2  (a^2+M N)} \left(  \mathbb{D}_0 \mu^z_\t -\mathbb{D}_\t \mu^z_0   \right) 
            \nn \\
   \o^{0 \t}_{\bZ}      &= \frac{-1 }{2 (a^2+M N)} \left(   a \mathcal{D}_\bz N -a \mathcal{D}_\bz M- N\mathcal{D}_\bz a + M\mathcal{D}_\bz a  \right) - \frac{t^2 \gamma_{z\bz}\exp(\Phi ) }{4 (a^2+M N)} \left(  \mathbb{D}_\t \mu^z_0  -\mathbb{D}_0 \mu^z_\t \right)
             \nn \\
     \o^{0 \t}_Z      &= \frac{ -1 }{ 2\left(a^2 + M N\right)}  \left(   a \mathcal{D}_z N -a  \mathcal{D}_z M- N \mathcal{D}_z a +M  \mathcal{D}_z a  \right)
     - \frac{ t^2 \gamma_{z\bz} \exp(\Phi ) }{4 \left(a^2 + M N\right)}  \left(\mathbb{D}_\t \mu^\bz_0 -\mathbb{D}_0 \mu^\bz_\t  \right) 
              \nn \\
   \o^{\t 0}_0          &= -\frac{a}{a^2+M N} \left(     \mathcal{D}_0 M+ \mathcal{D}_\t a  \right)  - \frac{N}{a^2 + MN}  \left( \mathcal{D}_0 a -\mathcal{D}_\t N  \right)
               \nn \\
    \o^{\t 0}_\t         &= -\frac{   a }{a^2+MN} \left(  \mathcal{D}_0 a- \mathcal{D}_\t N  \right) + \frac{M}{a^2+MN} \left( \mathcal{D}_0 M +\mathcal{D}_\t a  \right).
\end{align}
\restoregeometry

The action on all fields of the derivation  operation $\DD= \pa+... $ that figures in \eqref{Beltrami_sc} involves the derivatives  $\mathcal{D}_z, \mathcal{D}_\bz$ defined in \eqref{derD}. To make it explicit,  one must look at the details of the twelve independent $d=4$ vanishing  torsion conditions for $T^z=T^\bz=0$  in \eqref{Tz} and \eqref{Tbz}.  That is  
\begin{equation}
\label{D.14}
\begin{split}
\mathbb{D}_z \vpb &= \mathcal{D}_z \vpb - \frac{ \bp\mb} \mmb \\
\mathbb{D}_{\bar{z}} \vp &= \mathcal{D}_{\bar{z}} \vp - \frac{ \pa_z\m} \mmb 
\\
\mathbb{D}_z \Phi &= \demi \mathcal{D}_z \Phi - \frac{ \bp\mb} \mmb
\\
\mathbb{D}_{\bar{z}} \Phi &= \demi \mathcal{D}_{\bar{z}} \Phi - \frac{ \pa_z\m} \mmb 
\\
\mathbb{D}_z \mo &= \frac 1 \mmb  (  \pa_z\mo - \mob \pa_z\m + \mb (\pa_0 \m- \bp \mo))  \\
\mathbb{D}_{\bar{z}} \mob &= \frac 1 \mmb  (  \bp\mob - \mo \bp\mb + \m (\pa_0 \mb- \pa_z \mob))   \\
\mathbb{D}_0 \m &= \frac 1\mmb(  \pa_0 \m - \bp \mo - \mo\pa_z\m + \m\pa_z\mo)  \\
\mathbb{D}_0 \mb &= \frac 1\mmb(  \pa_0 \mb - \pa_z \mob - \mob\bp\mb + \mb\bp\mob)   \\
\mathbb{D}_z \mut &=  \frac 1 \mmb (  \pa_z\mut - \mubt \pa_z\m +  \mb (\partial_{\t} \m- \bp \mu^z_\t))  \\  
\mathbb{D}_{\bar{z}} \mu^\bz_\t &= \frac 1 \mmb (  {\color{black} \bp}\mubt - \mut \bp\mb +  \m (\pa_\t \mb- \pa_z \mu^\bz_\t) )   \\  
\mathbb{D}_\t \m &=   \frac 1 \mmb ( \pa_\t \m - \bp \mu^z_\t  - \mut  \pa_z\m  + \m\pa_z\mut)  \\
\mathbb{D}_\t \mb &=  \frac 1 \mmb ( \pa_\t \mb - \pa_z \mu^\bz_\t  - \mubt  \bp\mb  + \mb\bp\mubt)   \\
\mathbb{D}_\t \mo &= \pa_\t \mu^z_0 -  \frac{  1} {\mmb}   \big[ (\pa_z\m)\mo\mu^\bz_\t  + \bp\mo (\mu^\bz_\t  -\mb \mu^z_\t ) + \pa_z\mo(\mu^z_\t -\m\mu^\bz_\t )  + \pa_\t \m(\mbo-\mb \mo)          \big] \\
\mathbb{D}_\t \mbo &= \pa_\t \mbo - \frac{  1} {\mmb}  \big[ (\bp \mb) \mu^z_\t \mbo   + \bp \mbo (\mu^\bz_\t - \mb\mu^z_\t) + \pa_z \mbo (\mu^z_\t - \m \mu^\bz_\t) + \pa_\t \mb (\mo - \m \mbo) \big] \\
\mathbb{D}_0 \mu^z_\t &=  \pa_0 \mu ^z_\t - \frac{  1} {\mmb}  \big[ (\pa_z\m)\mu^z_\t   \mu^\bz_0   +\bp\mu^z_\t (\mbo-\mb \mo)  +  \pa_z\mu^z_\t (\mo-\m\mbo) + \pa_0\m (\mu^\bz_\t  -\mb \mu^z_\t )   \big] \\
\mathbb{D}_0 \mu^\bz_\t &= \pa_0 \mu^\bz_\t - \frac{  1} {\mmb}  \big[ (\bp \mb) \mo \mu^\bz_\t  +\bp \mu^\bz_\t (\mbo - \mb\mo) + \pa_z \mu^\bz_\t (\mo - \m \mbo ) + \pa_0 \mb (\mu^z_\t - \m \mu^\bz_\t)  \big].
\end{split}
\end{equation}

\section{$\Diff_4$    BRST Symmetry   
in the Beltrami Parametrization
}
\label{Annexe_D}
Here one computes the BRST $s$-transformations of all Beltrami fields and reparametrization ghosts.
They follow from the  ghost dependent  horizontality condition on the torsion that  encodes all the symmetries of the theory  \cite{p_forms}
\begin{equation}
\label{topological_torsion}
\tilde{T}^a \equiv  (d+s) \tilde e^a+(\omega^{a}_{\ b }  + \Omega^{a}_{\ b} ) \w   \tilde e^b=  \exp i_\xi \ T^a = 0
\end{equation}
where  $\xi^\mu(x) $ is the anticommuting reparametrization ghost and 
 \begin{align}
 \tilde e^z &=\exp i_\xi  e^z \equiv \exp \vp\  \tilde  \Et ^z , \nn \\
  \tilde e^\bz &=\exp i_\xi  e^\bz \equiv  \exp \vpb\  \tilde  \Et ^\bz , \nn\\
  \tilde e^0 &\equiv  N dt + a d\t + c^0 = N( dt + \xi^0 )  + a (d\t + \xi^\t) , \nn \\
  \tilde e^\t &\equiv  \bar{a} dt + M d\t + c^\t = \bar{a} (dt + \xi^0) + M (d\t + \xi^\t)
 \end{align}
 with  $\tilde{\Et}^z = dz + \m d\bz + \mo dt + \mt d\t +c^z$ and $\tilde{\Et}^\bz = d\bz + \mb dz + \mbo dt + \mbt d\t +c^\bz$. 
 The   ghost number $0$ components  of \eqref{topological_torsion}    lead to the $24$ linear equations that have already been solved in Appendix[\ref{Annexe_A}] to compute  the Beltrami  spin connection $\o(e)$.
  Its         ghost number $1$ components are  to determine the explicit expression of the anticommuting Lorentz ghosts  $\Omega^{ab} (x)$ in terms of the Beltrami fields and the reparametrization ghosts. 

  \noindent {\bf {    $\bullet$  Consequences of  $\tilde T^\t=  0
$}}
   \begin{align}
  0= \tilde  T^\t     
&\equiv     (d+s)  (e^\t +c^\t)    +  \demi t^2 \gamma_{z\bz} (\o^{\t z}  + \O^{\t z}) \w \tilde{e}^\bz +  \demi t^2 \gamma_{z\bz} ( \o^{\t \bz}  + \O^{\t \bz} ) \w \tilde{e}^z 
\nn  \\
& \ \ \ + ( \o^{\t 0}   + \O^{\t 0}  )\w (e^0  +c^0)
\nn \\
 &= de^\t + dz \pa_z c^\t + d\bz \bp c^\t + dt \pa_0 c^\t + d\t \pa_\t c^\t + sc^\t + (s\bar{a})dt +(sM)d\t 
 \nn   \\
&\ \ \  -    \demi t^2 \gamma_{z\bz}  {\exp \vpb } \  ( d\bz + \mb dz + \mbo dt + \mbt d\t +c^\bz) \w  ( \o^{\t z }    + \O^{\t z }   )
 \nn \\
  & \ \ \ -       \demi t^2 \gamma_{z\bz}  {\exp \vp } \ ( dz + \m d\bz + \mo dt + \mt d\t +c^z)  \w     (   \o^{\t \bz }    +   \O^{\t \bz }   )
  \nn \\
  &\ \ \ +  (   \o^{\t0}    +   \O^{\t0} )  \w (  N dt +a d\t    +c^0).
\end{align} 
The ghost number zero and form degree two  component  of this equation     is
\begin{equation}
\label{B2}
T^\t = 0 = de^\t + \demi t^2 \gamma_{z\bz}   \o^{\t z} \w e^\bz  + \demi t^2 \gamma_{z\bz}   \o^{\t \bz} \w e^z + \o^{\t 0} \w e^0
\end{equation}
which is  the  $\t$ component of the vanishing torsion conditions of   four-dimensional gravity    solved in Appendix[\ref{Annexe_A}].   Combined with  the other  ghost number zero components  of
$\tilde T^a=  0$ it 
   leads to the full system exposed and solved in Appendix[\ref{Annexe_A}], so one won't discuss the consequences of these ghost number zero equations in this appendix.

At ghost number one,     $\tilde T^\t=  0$ determines  the transformation laws of $M$ and $\bar{a}$ from its  components  proportional to $d\t$ and $dt$ as
\begin{align}
\label{sM}
sM &=   \pa_\t c^\t +  \demi t^2 \gamma_{z\bz}  \exp(\vpb) \ \Big(- \mbt \O^{\t z} + c^\bz \big(\o^{\t z}_\t + \mt \o^{\t z}_Z + \mbt \o^{\t z}_{\bZ} \big) \Big) 
\nn \\
&\ \ \ -  \demi t^2 \gamma_{z\bz}   \exp(\vp) \ \Big( \mu^z_\t \O^{\t \bz} - c^z \big( \o^{\t \bz}_{\t} + \mt \o^{\t \bz}_Z + \mbt \o^{\t \bz}_{\bZ}   \big) \Big) + c^0 \Big(  \o^{\t 0}_\t +\mt \o^{\t 0}_Z + \mbt \o^{\t 0}_{\bZ}   \Big) - a \O^{\t 0},
\\
\label{sba}
s\bar{a} &=   \pa_0 c^\t +  \demi t^2 \gamma_{z\bz}   \exp(\vpb) \ \Big( - \mbo \O^{\t z} + c^\bz \big(\o^{\t z}_0 + \mo \o^{\t z}_Z + \mbo \o^{\t z}_{\bZ} \big) \Big) 
\nn \\
&\ \ \ -  \demi t^2 \gamma_{z\bz}   \exp(\vp) \ \Big( \mu^z_0 \O^{\t \bz} - c^z \big(\o^{\t \bz}_0 + \mo \o^{\t \bz}_Z + \mbo \o^{\t \bz}_{\bZ}   \big) \Big)  +  c^0 \Big(   \o^{\t 0}_0 + \mo \o^{\t 0}_Z + \mbo \o^{\t 0}_{\bZ}     \Big)   - N \O^{\t 0}.
\end{align}
The terms proportional to $dz$ and $d\bz$ with ghost number one imply  the  following constraints for $\O^{\t z}$ and $\O^{\t \bz}$
 \begin{align}
 \label{C.6}
 \demi t^2 \gamma_{z\bz}   \Big(    \exp(\vp)  \O^{\t \bz} + \mb \exp(\vpb)  \O^{\t z} \Big) &=  \pa_z c^\t + \demi t^2 \gamma_{z\bz}  \Big(  \exp(\vp)(\o^{\t \bz}_Z + \mb \o^{\t \bz}_{\bZ})c^z 
 \nn \\ 
 & \ \ \ + \exp(\vpb)(\o^{\t z}_Z + \mb \o^{\t z}_{\bZ})c^\bz   \Big)  + (\o^{\t 0}_Z + \mb \o^{\t 0}_{\bZ})c^0  ,
\\
  \label{C.7}
  \demi t^2 \gamma_{z\bz}   \Big(  \m  \exp(\vp)  \O^{\t \bz} +  \exp(\vpb)  \O^{\t z} \Big) &=   \bp c^\t +\demi t^2 \gamma_{z\bz}  \Big(  \exp(\vp)(\o^{\t \bz}_{\bZ} + \m \o^{\t \bz}_Z)c^z  \nn \\
 & \ \ \  + \exp(\vpb)(\o^{\t z}_{\bZ} + \m \o^{\t z}_Z)c^\bz   \Big) + (\o^{\t 0}_{\bZ} + \m \o^{\t 0}_Z)c^0.
 \end{align}
Solving  this system of two equations  determines the Lorentz ghosts $\O^{\t z}$ and $\O^{\t \bz}$, in terms of the Beltrami reparametrization ghosts, that have to be plugged back in (\ref{sM}), (\ref{sba}) and (\ref{sct}).  The $\o^{ab}_\mu $ are the Beltrami field dependent  expressions determined in Appendix[\ref{Annexe_A}].
At ghost number two,   $\tilde T^\t=0$  determines  the transformation law of the reparametrization ghost $c^\t$ as
\begin{equation}
\label{sct}
sc^\t =   \demi t^2 \gamma_{z\bz}   \Big(  \exp(\vpb) c^\bz \O^{\t z} + \exp(\vp)  c^z \O^{\t \bz}     \Big)  - \O^{\t 0} c^0.
\end{equation}

 \noindent {\bf {    $\bullet$  Consequences of  $\tilde T^0=   0
$}}
\begin{align}
0 = \tilde  T^0 &\equiv     (d+s)  (e^0 +c^0)    -  \demi t^2 \gamma_{z\bz}   (\o^{z0}  + \O^{z0} ) \w \tilde{e}^\bz +  \demi t^2 \gamma_{z\bz}   ( \o^{0\bz}  + \O^{0\bz} ) \w \tilde{e}^z 
\nn \\
&\ \ \ - ( \o^{0 \t}   + \O^{0 \t}  )\w (e^\t  +c^\t)
\nn  \\
 &= de^0 + dz \pa_z c^0 + d\bz \bp c^0 + dt \pa_0 c^0 + d\t \pa_\t c^0 + sc^0 + (sN)dt +(sa)d\t 
  \nn  \\
&\ \ \  +   \demi t^2 \gamma_{z\bz}  {\exp \vpb } \  ( d\bz + \mb dz + \mbo dt + \mbt d\t +c^\bz) \w  ( \o^{z0}    + \O^{z0}   )
\nn  \\
  & \ \ \ -       \demi t^2 \gamma_{z\bz}   {\exp \vp } \ ( dz + \m d\bz + \mo dt + \mt d\t +c^z)  \w     (   \o^{0\bz}    +   \O^{0\bz}   ) 
  \nn \\
  &\ \ \ -  (   \o^{0 \t}    +   \O^{0 \t} )  \w (  \bar{a} dt +M d\t    +c^\t).
\end{align} 
At ghost number one, the terms proportional to $dt$ and $d\t$ provide
\begin{align}
\label{sN}
sN &=  \pa_0 c^0 +\demi t^2 \gamma_{z\bz}  \exp(\vpb) \ \Big( \mbo \O^{z0} - c^\bz \big(\o^{z0}_0 + \mo \o^{z0}_Z + \mbo \o^{z0}_{\bZ}\big) \Big) 
\nn \\
&\ \ \ - \demi t^2 \gamma_{z\bz} \exp(\vp) \ \Big( \mu^z_0 \O^{0\bz} - c^z \big(\o^{0\bz}_0 + \mo \o^{0\bz}_Z + \mbo \o^{0\bz}_{\bZ}   \big) \Big) -c^\t \Big(  \o^{0 \t}_0 +\mo \o^{0 \t}_Z + \mbo \o^{0 \t}_{\bZ}   \Big)+ \bar{a} \O^{0 \t},
\\
\label{sa}
sa &=   \pa_\t c^0 + \demi t^2 \gamma_{z\bz}  \exp(\vpb) \ \Big( \mbt \O^{z0} - c^\bz \big(\o^{z0}_\t + \mt \o^{z0}_Z + \mbt \o^{z0}_{\bZ} \big) \Big) 
\nn \\
&\ \ \  - \demi t^2 \gamma_{z\bz}  \exp(\vp) \ \Big( \mu^z_\t \O^{0\bz} - c^z \big(\o^{0\bz}_\t + \mt \o^{0\bz}_Z + \mbt \o^{0\bz}_{\bZ}   \big) \Big) -c^\t \Big(   \o^{0 \t}_\t + \mt \o^{0 \t}_Z + \mbt \o^{0 \t}_{\bZ}     \Big)   + M \O^{0 \t}.
\end{align}
Still at  ghost number one, the terms proportional to $dz$ and $d\bz$ imply that   $\O^{z0}$ and $\O^{0\bz}$   satisfy 
 \begin{align}
 \label{C.13}
\demi t^2 \gamma_{z\bz}  \Big(    \exp(\vp)  \O^{0\bz} - \mb \exp(\vpb)  \O^{z0} \Big) &=  \demi t^2 \gamma_{z\bz}  \Big(  \exp(\vp)(\o^{0\bz}_Z + \mb \o^{0\bz}_{\bZ})c^z - \exp(\vpb)(\o^{z0}_Z + \mb \o^{z0}_{\bZ})c^\bz   \Big)
 \nn \\
 & \ \ \   + \pa_z c^0 - (\o^{0 \t}_Z + \mb \o^{ 0 \t}_{\bZ})c^\t 
\\
  \label{C.14}
 \demi t^2 \gamma_{z\bz}  \Big(  \m  \exp(\vp)  \O^{0\bz} -  \exp(\vpb)  \O^{z0} \Big) &=    \demi t^2 \gamma_{z\bz}  \Big(  \exp(\vp)(\o^{0\bz}_{\bZ} + \m \o^{0\bz}_Z)c^z - \exp(\vpb)(\o^{z0}_{\bZ} + \m \o^{z0}_Z)c^\bz   \Big) 
 \nn \\
 &\ \ \ + \bp c^0 - (\o^{0 \t }_{\bZ} + \m \o^{0 \t }_Z)c^\t .
 \end{align}
This system leads to the determination of $\O^{z0}$ and $\O^{0\bz}$ that have to be    plugged back in \eqref{sN}, \eqref{sa} and \eqref{sc0}. Then, by combining  \eqref{sba} and \eqref{sa},  one  can express  $\O^{0 \t}$ in terms of $s(a + \bar{a})$.
 
The  ghost    number $2$ components of the equation   $\tilde T^0=0$ imply
\begin{equation}
\label{sc0}
sc^0 = \demi t^2 \gamma_{z\bz}  \Big(  \exp(\vp)  c^z \O^{0\bz} -  \exp(\vpb) c^\bz \O^{z0}    \Big)  + \O^{0 \t} c^\t .
\end{equation}

 \noindent {\bf {    $\bullet$ Consequences of   $\tilde T^z=  0 $    } }
\begin{align}
0 =  \tilde T^z    
  &\equiv  
  ( d + s) \tilde{e}^z - \demi t^2 \gamma_{z\bz}  (\o^{{\bz z}} + \O^{{\bz z}}) \w \tilde{e}^z + (\o^{z0} + \O^{z0}) \w (e^0 + c^0) - (\o^{z \t} + \O^{z \t} ) \w (e^\t + c^\t) 
  \nn \\
  &= \exp(\vp)   \Big[  \big( (d+s)\vp - \demi t^2 \gamma_{z\bz} (\o^{{\bz z}} + \O^{{\bz z}}) \big) \w \tilde{\mathcal{E}}^z + (d+s) (  \m d\bz + \mo dt + \mt d\t + c^z ) \Big]
  \nn\\
  &\ \ \  -(  Ndt +ad\t + c^0  ) \w (  \o^{z0} + \O^{z0}  )   +  (   \bar{a}dt + Md\t + c^\t  ) \w  (   \o^{z \t} + \O^{z \t}  ) .
\end{align}
The terms at ghost number one and proportional to $dz$   imply
\begin{equation}
\label{svp}
s\vp =  \pa_z c^z +  c^z  \Big(  \pa_z \vp -    \frac{t^2}{2} \gamma_{z\bz} (  \o^{{\bz z}}_Z + \mb \o^{{\bz z}}_{\bZ})  \Big)
+ \frac{t^2}{2} \gamma_{z\bz}  \O^{{\bz z}}  +\exp(-\vp) \Big[  c^0 \big(  \o^{z0}_Z + \mb \o^{z0}_{\bZ}   \big) - c^\t  \big(  \o^{z \t}_Z + \mb \o^{z \t}_{\bZ}  \big)   \Big]  .
\end{equation}
 Once combined with the  analogous equation for $s\vpb$ implied by $\tilde T^\bz=  0 $, this equation  will   determine    $\O^{{\bz z}}$ as a function of the Beltrami fields and reparametrization ghosts, as well as the expressions of 
  $s\vp $ and  $s\vpb$.  

It is then  useful to multiply the BRST equation $\tilde T^z=  0 $   by $\tilde{\mathcal{E}}^z$. Indeed, since $\tilde{\mathcal{E}}^z \w \tilde{\mathcal{E}}^z =0$, one gets  the simpler equation
\begin{align}
\label{magicz}
0 &= \big( dz + \m d\bz + \mo dt + \mt d\t + c^z \big) \big( d+s \big) \big( \m d\bz + \mo dt + \mt d\t + c^z \big) 
\nn \\
&  \ \ \  -\exp(-\vp) \big( dz + \m d\bz + \mo dt + \mt d\t + c^z \big) \w \big( Ndt + a d\t + c^0 \big) \w \big( \o^{z0} + \O^{z0} \big) 
\nn \\
&  \ \ \ +\exp(-\vp) \big( dz + \m d\bz + \mo dt + \mt d\t + c^z \big) \w \big( \bar{a}dt + Md\t + c^\t \big) \w \big( \o^{z \t} + \O^{z \t} \big) .
\end{align} 
At ghost number one, the terms proportional to $dz \w d\bz$, $dz \w dt$ and $dz \w d\t$  stemming  from  this equation give respectively the transformation laws $s\m$, $s\mo$ and $s\mt$, that is 
 \begin{align}
 \label{magiczz}
s \m &=  \bp c^z + c^z \pa_z \m - \m \pa_z c^z+ \exp(-\vp) \ \Big[  c^0 (\mmb)\o^{z0}_{\bZ} -   c^\t (\mmb) \o^{z \t}_{\bZ} \Big],
\\
s \mo &=  \pa_0 c^z + c^z \pa_z \mo - \mo \pa_z c^z - \exp(-\vp) \Big[ N \big( \O^{z0} - c^z(\o^{z0}_Z + \mb \o^{z0}_{\bZ}) \big) - c^0 \big( \o^{z0}_0 + \o^{z0}_{\bZ} (  \mbo - \mb \mo ) \big) \Big] \nn
\\
&\ \ \ + \exp(-\vp)  \Big[ \bar{a} \big( \O^{z \t} - c^z ( \o^{z \t}_Z + \mb \o^{z \t}_{\bZ} ) \big) - c^\t \big( \o^{z \t}_0 + \o^{z \t}_{\bZ} (\mbo - \mb \mo) \big) \Big],
\\
s \mt &=  \pa_\t c^z + c^z \pa_z \mt - \mt \pa_z c^z - \exp(-\vp) \Big[ a \big( \O^{z0} - c^z(\o^{z0}_Z + \mb \o^{z0}_{\bZ}) \big) - c^0 \big( \o^{z0}_\t + \o^{z0}_{\bZ} (  \mbt - \mb \mt ) \big) \Big]
\nn
\\
&\ \ \ + \exp(-\vp)  \Big[ M \big( \O^{z \t} - c^z ( \o^{z \t}_Z + \mb \o^{z \t}_{\bZ} ) \big) - c^\t \big( \o^{z \t}_\t + \o^{z \t}_{\bZ} (\mbt - \mb \mt) \big) \Big],
\end{align}
while  the terms  with ghost number $2$ proportional to $dz$ imply 
 \begin{equation}
 sc^z =  c^z \pa_z c^z - \exp(-\vp) \big( \O^{z0} - c^z (\o^{z0}_Z + \mb \o^{z0}_{\bZ}) \big) c^0
+ \exp(-\vp) \big( \O^{z \t} -c^z(\o^{z \t}_Z + \mb \o^{z \t}_{\bZ} ) \big) c^\t .
\end{equation}
 
 \noindent {\bf {    $\bullet$ Consequences of   $\tilde T^\bz=  0 $    } } 
    \begin{align}
    \label{tTbz}
0 = \tilde T^\bz    
  &\equiv ( d + s) \tilde{e}^\bz + \demi t^2 \gamma_{z\bz}  (\o^{{\bz z}} + \O^{{\bz z}}) \w \tilde{e}^\bz - (\o^{0\bz} + \O^{0\bz}) \w (e^0 + c^0) - (\o^{\bz \t} + \O^{\bz \t} ) \w (e^\t + c^\t) \nn \\
  &=  \exp(\vpb)    \Big[ \big( (d+s)\vpb + \demi t^2 \gamma_{z\bz} (\o^{{\bz z}} + \O^{{\bz z}}) \big) \w \tilde{\mathcal{E}}^\bz + (d+s)(  \mb dz + \mob dt + \mbt d\t + c^\bz   ) \Big]
  \nn\\
  & \ \ \  + (   Ndt +ad\t + c^0  ) \w (  \o^{0\bz} + \O^{0\bz}  )   +  (   \bar{a}dt + Md\t + c^\t  ) \w  (   \o^{\bz \t} + \O^{\bz \t}  )  .
\end{align}
The term at ghost number one and proportional to $d\bz$ is the analog of  \eqref{svp}, that is  
\begin{equation}
\label{svpb}
s\vpb =  \bp c^\bz +  c^\bz  \Big( \bp \vpb   +\frac{t^2}{2} \gamma_{z\bz} (\o^{{\bz z}}_{\bZ} + \m \o^{{\bz z}}_Z) \Big)  - \frac{t^2}{2} \gamma_{z\bz} \O^{{\bz z}}  - \exp(-\vpb)\ \Big[ c^0 \big(\o^{0\bz}_{\bZ} + \m \o^{0\bz}_Z \big) + c^\t \big( \o^{\bz \t}_{\bZ} + \m \o^{\bz \t}_Z \big)  \Big].
\end{equation}  
One can   then   combine    \eqref{svp} and \eqref{svpb} and determine  $\O^{{\bz z}}$ as a function of $s(\vp - \vpb)$ and   the reparametrization ghosts.  
This  expression, as well  as those  for $\O^{0 \t}$, $\O^{z0}$, $\O^{0\bz}$, $\O^{\t z}$ and $\O^{\t \bz}$, is displayed in \eqref{C.30}.

 By multiplying   (\ref{tTbz}) by $\tilde{\mathcal{E}}^\bz$, one gets the analog of \eqref{magicz}:
\begin{align}
\label{magiczb}
0 &= \big( d\bz + \mb dz + \mob dt + \mbt d\t + c^\bz \big) \big( d+s \big) \big( \mb dz + \mob dt + \mbt d\t + c^\bz \big) 
\nn\\
&\ \ \  + \exp(-\vpb) \big( d\bz + \mb dz + \mob dt + \mbt d\t + c^\bz \big) \w \big( Ndt + a d\t + c^0 \big) \w \big( \o^{0\bz} + \O^{0\bz} \big) 
\nn\\
&\ \ \  + \exp(-\vpb) \big( d\bz + \mb dz + \mob dt + \mbt d\t + c^\bz \big) \w \big( \bar{a}dt + Md\t + c^\t \big) \w \big( \o^{\bz \t} + \O^{\bz \t} \big),
\end{align}
which provides 
\begin{align}
\label{magiczzb}
 s \mb &=  \pa_z c^\bz + c^\bz \bp \mb - \mb \bp c^\bz- \exp(-\vpb) \ \Big[  c^0 (\mmb)\o^{0\bz}_Z +   c^\t (\mmb) \o^{\bz \t}_Z \Big] ,
\\
s \mob &=   \pa_0 c^\bz + c^\bz \bp \mob - \mob \bp c^\bz + \exp(-\vpb) \Big[ N \big( \O^{0\bz} - c^\bz(\o^{0\bz}_{\bZ} + \m \o^{0\bz}_Z) \big) - c^0 \big( \o^{0\bz}_0 + \o^{0\bz}_Z (  \mo - \m \mob ) \big) \Big]
\nn\\
&\ \ \  + \exp(-\vpb)  \Big[ \bar{a} \big( \O^{\bz \t} - c^\bz ( \o^{\bz \t}_{\bZ} + \m \o^{\bz \t}_Z ) \big) - c^\t \big( \o^{\bz \t}_0 + \o^{\bz \t}_Z (\mo - \m \mob) \big) \Big]  ,
\\
s \mbt &=  \pa_\t c^\bz + c^\bz \bp \mbt - \mbt \bp c^\bz + \exp(-\vpb) \Big[ a \big( \O^{0\bz} - c^\bz(\o^{0\bz}_{\bZ} + \m \o^{0\bz}_Z) \big) - c^0 \big( \o^{0\bz}_\t + \o^{0\bz}_Z (  \mt - \m \mbt ) \big) \Big]
\nn\\
&\ \ \ + \exp(-\vpb)  \Big[ M \big( \O^{\bz \t} - c^\bz ( \o^{\bz \t}_{\bZ} + \m \o^{\bz \t}_Z ) \big) - c^\t \big( \o^{\bz \t}_\t + \o^{\bz \t}_Z (\mt - \m \mbt) \big) \Big]
\end{align}
and
\begin{equation}
 sc^\bz =  c^\bz \bp c^\bz + \exp(-\vpb) \big( \O^{0\bz} - c^\bz (\o^{0\bz}_{\bZ} + \m \o^{0\bz}_Z) \big) c^0 +
 \exp(-\vpb) \big( \O^{\bz \t} -c^\bz(\o^{\bz \t}_{\bZ} + \m \o^{\bz \t}_Z ) \big) c^\t .
 \end{equation}

Let us now display the expressions of the $6$ Lorentz ghosts that have been determined in functions of the Beltrami fields  and the Beltrami reparametrization ghosts in this appendix, that is:
\begin{align}
\label{C.30}
  \O^{{\bz z}} &=   \frac{c^\bz}{2} ( \o^{{\bz z}}_{\bZ} + \m \o^{{\bz z}}_Z)   + \frac{c^z}{2} (\o^{{\bz z}}_Z + \mb \o^{{\bz z}}_{\bZ}  ) +  \frac{1}{t^2 \gamma_{z\bz}}  \bigg\{   s(\vp - \vpb)   + \big( \bp c^\bz + (\bp \vpb)c^\bz \big)
  \nn \\    
 &\ \ \   -\big( \pa_z c^z  + (\pa_z \vp) c^z \big) -  \exp(-\vpb)\ \Big[ c^0 \big(\o^{0\bz}_{\bZ} + \m \o^{0\bz}_Z \big) + c^\t \big( \o^{\bz \t}_{\bZ} + \m \o^{\bz \t}_Z \big)  \Big] 
 \nn \\
 &\ \ \ - \exp(-\vp) \Big[  c^0 \big(  \o^{z0}_Z + \mb \o^{z0}_{\bZ}   \big) - c^\t  \big(  \o^{z \t}_Z + \mb \o^{z \t}_{\bZ}  \big)   \Big] \bigg\} ,
 \nn \\
 \O^{z0} &= -   \dfrac{2 \exp(-\vpb)}{t^2 \gamma_{z\bz}(\mmb)} \Big[   \bp c^0 - \m \pa_z c^0  \Big]  - \exp(\vp - \vpb)\  c^z \o^{0\bz}_{\bZ} +  c^\bz \o^{z0}_{\bZ} -   \frac{2}{t^2\gamma_{z\bz}}  \exp(-\vpb)\ c^\t \o^{\t 0}_{\bZ},
 \nn \\
 \O^{0\bz} &=  \dfrac{2 \exp(-\vp)}{t^2 \gamma_{z\bz}(\mmb)} \Big[  \pa_z c^0 - \mb \bp c^0  \Big]  + c^z \o^{0\bz}_Z - \exp(\vpb - \vp) \ c^\bz \o^{z0}_Z +  \frac{2}{t^2 \gamma_{z\bz}} \exp(-\vp)\ c^\t \o^{\t 0}_Z,
 \nn \\
 \O^{\t z} &=   \dfrac{2 \exp(-\vpb)}{t^2 \gamma_{z\bz} (\mmb)} \Big[   \bp c^\t - \m \pa_z c^\t      \Big]  + \exp(\vp - \vpb)\ c^z \o^{\t \bz}_{\bZ} +  c^\bz \o^{\t z}_{\bZ} +  \frac{2}{t^2\gamma_{z\bz}} \exp(-\vpb) \ c^0 \o^{\t 0}_{\bZ},
 \nn \\
 \O^{\t \bz} &= \dfrac{2 \exp(-\vp)}{t^2 \gamma_{z\bz} (\mmb)} \Big[  \pa_z c^\t - \mb \bp c^\t      \Big]  + c^z \o^{\t \bz}_Z + \exp(\vpb - \vp) \ c^\bz \o^{\t z}_Z +  \frac{2}{t^2 \gamma_{z\bz} }  \exp(-\vp) \ c^0 \o^{\t 0}_Z,
 \nn \\
 \O^{\t 0} &= \frac{1}{(N + M) } \bigg\{ -s (a + \bar{a} )  +   \pa_\t c^0 + \pa_0 c^\t + c^0 \o^{\t 0}_0 + c^\t \o^{\t 0}_\t + \demi t^2 \gamma_{z\bz} \Big[ \exp (\vpb) \ c^\bz(\o^{\t z}_0 - \o^{z 0}_\t ) 
 \nn \\
 &\ \ \ + \exp(\vp) \ c^z(\o^{\t \bz}_0  + \o^{0 \bz}_\t) \Big] - \frac{1}{\mmb} \Big[ \mu^\bz_\t (\bp c^0 - \m \pa_z c^0) +  \mu^z_\t (\pa_z c^0 - \mb \bp c^0) 
 \nn \\
 &\ \ \  + \mu^\bz_0 (\bp c^\t - \m \pa_z c^\t) + \mu^z_0 (\pa_z c^\t - \mb \bp c^\t)    \Big] \bigg\},
\end{align}
which are completely determined when $a = - \bar{a}$ and $\vp = \vpb = \frac{\Phi}{2}$, just like in the Beltrami parametrization. In this particular case, which is the one useful for this article,  the $s$ transformations of the Beltrami fields can be fully determined by plugging the exact expression of the spin connection $\o(e)$ \eqref{Beltrami_sc}, determined in Appendix[\ref{Annexe_A}], in the $s$ transformations just derived in this appendix.

\newgeometry{top=2cm,bottom=2.5cm}
Some important exact results in the context of the BMS symmetry are:
\begin{align}
\label{sm_exact}
s \m &= \bp c^z + c^z \pa_z \m - \m \pa_z c^z -\frac{ \mmb  }{a^2+M N}  \Big( -a c^0 \mathbb{D}_\t \m+a c^\t \mathbb{D}_0 \m-Mc^0 \mathbb{D}_0 \m -Nc^\t \mathbb{D}_\t \m  \Big),
\\
\label{smuzu_exact}
2s \mu^z_u &= s (\mt - \mo)
\nn \\
&= \pa_\t c^z - \pa_0 c^z +2 ( c^z  \pa_z \mu^z_u - \mu^z_u \pa_z c^z) +(c^\bz-\mb c^z )  (\mathbb{D}_0 \m-\mathbb{D}_\t \m) 
\nn \\
&\ \ \ + \frac{2\exp(-\Phi )}{t^2 \gamma_{z\bz} (\mmb)} \Big(
 (a-N) ( \bp c^0  -\m \pa_z c^0  )  - (a+M) (\bp c^\t  - \m \pa_z c^\t)  
\Big)
\nn \\
&\ \ \ + \frac{1}{(a^2 +MN)} \Big( \big[ (a+M) c^0 - (a-N)c^\t \big] \big(\mathbb{D}_0 \mt-\mathbb{D}_\t \mo \big) 
\nn \\
&\ \ \ +2 (\mu^\bz_u-\mb \mu^z_u) \big[a c^0  \mathbb{D}_\t \m  -a c^\t \mathbb{D}_0 \m +  M c^0  \mathbb{D}_0 \m  +Nc^\t  \mathbb{D}_\t \m \big]  \Big)
\nn \\
&\ \ \ \frac{-\exp(- \Phi ) }{t^2 \gamma_{z\bz} (a^2+M N)}
\Bigg(
2a^2 
\big[
 c^0 (\mathcal{D}_\bz a-\mathcal{D}_\bz M)
- c^\t (\mathcal{D}_\bz a+\mathcal{D}_\bz N)
\big]
\nn \\
&\ \ \ 
+2 a \big[ Mc^0  (\mathcal{D}_\bz a-\mathcal{D}_\bz M+\mathcal{D}_\bz N)
+Nc^\t  (\mathcal{D}_\bz a-\mathcal{D}_\bz M+\mathcal{D}_\bz N)-Nc^0 \mathcal{D}_\bz a -Mc^\t \mathcal{D}_\bz a  \big]
\nn \\
&\ \ \ +2 \big[ M^2 c^0 \mathcal{D}_\bz a - MN c^0 \mathcal{D}_\bz N - N^2 c^\t \mathcal{D}_\bz a - MN c^\t \mathcal{D}_\bz M   \big]
  \Bigg) ,
\\
\label{sM_exact}
2 s \mathcal{M} &= s(M+N)
\nn \\
&= \frac{1}{\mathcal{M}^2}  \Big( a^2 (c^\bz (\mathcal{D}_\bz M+\mathcal{D}_\bz N)+c^z (\mathcal{D}_z M+\mathcal{D}_z N)+\pa_0 c^0)+\pa_\t c^\t \left(a^2+M N\right)
\nn \\
&\ \ \ -a (c^0 (\mathcal{D}_0 a-\mathcal{D}_\t N)+c^\t (\mathcal{D}_0 M+\mathcal{D}_\t a))+c^0 M (\mathcal{D}_0 M+\mathcal{D}_\t a)+M N \big(c^\bz (\mathcal{D}_\bz M+\mathcal{D}_\bz N)
\nn \\
&\ \ \ +c^z (\mathcal{D}_z M+\mathcal{D}_z N)+\pa_0 c^0 \big) +c^\t N (\mathcal{D}_\t N-\mathcal{D}_0 a) \Big) 
\nn \\
&\ \ \  -  \frac{1}{\mmb} \Big( \bp c^0 (\mob-\mb \mo)+\bp c^\t (\mbt-\mb \mt) +\pa_z c^0 (\mo-\mob \m)+\pa_z c^\t (\mt-\mbt \m) \Big) ,
\\
\label{scz_exact}
sc^z &= c^z \pa_z c^z + \frac{ 1}{\mathcal{M}^2} (c^\bz-c^z \mb) (-a c^0 \mathbb{D}_\t \m+a c^\t \mathbb{D}_0 \m- Mc^0 \mathbb{D}_0 \m - Nc^\t \mathbb{D}_\t \m )
\nn \\
&\ \ \ +\frac{2 \exp(-\Phi )}{t^2 \gamma_{z\bz} (\mmb)} (-c^0 \bp c^0+c^\t \bp c^\t+c^0 \m \pa_z c^0-c^\t \m \pa_z c^\t) ,
\\
\label{scu_exact}
sc^u &= s (c^\t - c^0)
\nn \\
&= -\frac{1}{4 (M+N)}    \Bigg( \frac{2} {a^2+M N}     
\Big[  
-2 
\Big\{
 a^2 \pa_0 c^\t  +MN \pa_0 c^\t   
 \nn \\
 &\ \ \ +N c^0 (\mathcal{D}_\t N-\mathcal{D}_0 a)
+M c^\t  (\mathcal{D}_0 M+\mathcal{D}_\t a) -a c^0 (\mathcal{D}_0 M+\mathcal{D}_\t a)+ac^\t (\mathcal{D}_0 a-\mathcal{D}_\t N)
\Big\} c^u 
 \nn \\
 &\ \ \ -t^2  \gamma_{z\bz} (M+N) \exp(\Phi )  
 \Big\{
 a     c^\bz c^z \big( \mathcal{D}_\t \Phi- \mathcal{D}_0 \Phi
+\mathbb{D}_\bz \mob+\mathbb{D}_z \mo - ( \mathbb{D}_\bz \mbt+\mathbb{D}_z \mt)
\big)
\nn \\
&\ \ \ +c^\bz c^z \big( M \mathcal{D}_0 \Phi +N\mathcal{D}_\t \Phi -M (\mathbb{D}_\bz \mob+\mathbb{D}_z \mo) 
-N (\mathbb{D}_\bz \mbt +\mathbb{D}_z \mt) \big) 
 \Big\}
 \Big]    
\nn \\
&\ \ \ -\frac{ 4 \bp c^0}{\mmb} \Big(      -c^u (\mu^\bz_\t-\mb \mu^z_\t) + (c^\bz - \mb c^z  ) (M+N) \Big)
\nn \\
&\ \ \ -\frac{4  \bp c^\t }{\mmb}   \Big(     -c^u (\mob-\mb \mo)  -   ( c^\bz  - \mb c^z  ) (M+N)  \Big)
\nn \\
&\ \ \ -\frac{4}{\mmb}    \Big( - \pa_z c^0 \big(   c^u(\mu^z_\t - \m\mu^\bz_\t )   -(c^z - \m c^\bz)  (M+N)  \big)
\nn \\
&\ \ \ -\pa_z c^\t \big(c^u (\mo-\m \mob )- (c^z - \m c^\bz)  (M+N) \big) +  \pa_\t c^0 (\mmb) c^u \Big)  \Bigg).
\end{align}
\restoregeometry

\newpage

\bibliography{biblio}
\bibliographystyle{kp}

\end{document}